\documentclass[11pt, oneside]{article}   	
\usepackage{geometry}                		
\title{Screened Coulomb Interactions With Non-uniform Surface Charge}
\author{Sandip Ghosal\\
Department of Mechanical Engineering\\
\& Engineering Sciences and Applied Mathematics,\\
Northwestern University, Evanston, IL 60208, USA\\[2ex]
John D. Sherwood\\
Department of Applied Mathematics and Theoretical Physics,\\
University of Cambridge, Cambridge CB3 0WA, UK
}
\date{\today}							
\usepackage{fancyhdr}
\usepackage{graphicx}
\usepackage{dcolumn}
\usepackage{bm}
\usepackage{epsfig}
\usepackage{amsmath}
\usepackage{amssymb}
\usepackage{psfrag}
\usepackage[square,sort&compress,comma,numbers]{natbib}
\usepackage{epstopdf}
\usepackage{color}
\DeclareMathOperator{\sech}{sech}
\DeclareMathOperator{\cosech}{cosech}

\begin{document}
\maketitle


\begin{abstract}
The screened Coulomb interaction between a pair of infinite parallel planes with spatially varying surface charge is considered in the limit of small electrical potentials for arbitrary Debye lengths. A simple expression for the disjoining pressure is derived in terms of a two dimensional integral in Fourier space. The integral is evaluated for periodic and random charge distributions and the disjoining pressure is expressed as a sum over Fourier-Bloch reciprocal lattice vectors  or in terms of an integral involving the autocorrelation function respectively. The force between planes with a finite area of uniform charge, a model for the DLVO interaction between finite surfaces, is also calculated. It is shown that the overspill of the charge cloud beyond the region immediately between the charged areas results in a reduction of the disjoining pressure, as reported by us recently in the long Debye length limit for planes of finite width.
\end{abstract}
\section{Introduction}
\label{sec_intro} 
The force of interaction between neighboring dielectric-electrolyte interfaces is responsible for 
a wide range of phenomena such as the stability of colloidal suspensions \cite{russel_colloidal_1989}, 
colloidal self-assembly \cite{zhang_toward_2015},
and the stability of thin liquid films \cite{danov_effect_2004}. The nature of the interaction is generally 
a short range
($\sim$ 1 nm)
attraction due to molecular van der Waals forces and a longer range ($\sim$ 100 nm) electrostatic 
repulsion that is partially screened by the free ions in the electrolyte. 
Analysis of the force between particles was provided by 
Derjaguin \& Landau \cite{derjaguin_theory_1941} 
and by 
Verwey and Overbeek \cite{verwey_theory_1999}.
The expression for the force or the interaction potential 
that they derived and the theory of colloid 
stability based on it has become the bedrock of most investigations of the subject and is referred to 
as ``DLVO theory''. The original DLVO approximation has been improved and 
extended \cite{mccartney_improvement_1969,bell_approximate_1970,
white_deryaguin_1983,bhattacharjee_surface_1997,schnitzer_generalized_2015} over the years. 
When DLVO theory was developed, it could only be tested by comparing its predictions 
against experimental observations
of large scale phenomena such as the onset of flocculation in colloidal systems.
This changed with the introduction of the surface force apparatus
capable of measuring pico-Newton interaction forces between atomically smooth mica surfaces \cite{tabor_surface_1968,tabor_recent_1969,white_dispersion_1976,israelachvili_measurement_1978,israelachvili_solvation_1987}. More recently, 
other precision instruments such as atomic force microscopes \cite{todd_probing_2004} and 
laser optical tweezers \cite{sugimoto_direct_1997,gutsche_forces_2007} have been used to 
measure directly these small scale interfacial forces. These experiments have generally 
confirmed the theory within its expected range of validity. However, uncertainties in the 
interpretation of these experiments remain for separations smaller than a few nanometers, in the 
regime where the interaction changes from repulsive to 
attractive \cite{israelachvili_international_1977,ninham_progress_1999}
due to the molecular van der Waals forces.

In addition to new and distinct mechanisms \cite{ninham_progress_1999} that have been discussed 
to explain the interaction at short distances, the details of the screened 
Coulomb interaction model itself are important.
For example, at such small separations, the predictions of the constant charge, constant potential 
and charge regulation models yield distinctly different results 
\cite{russel_colloidal_1989, ninham_electrostatic_1971,ninham_progress_1999,boon_charge_2011,
bell_calculation_1972}. Furthermore, in this 
regime, it is impossible to invoke the simplifying
assumption 
that there is only a slight overlap of the Debye layers adjacent to the two surfaces. If the lateral extent of the interacting surfaces is also small 
(as in nanocolloids, tips of atomic force microscopes or clay particles) then 
overspill of the Debye layer out of the region immediately between the surfaces
can become 
important \cite{ghosal_repulsion_2016}. 
Recent variants of the classical surface force experiment have shown 
significant anomalies that have not yet been fully explained: surfactant coated mica surfaces 
show a long range attraction \cite{christenson_direct_2001,meyer_origin_2005,perkin_long-range_2006} in place of the
expected repulsion.
It has been suggested \cite{perkin_long-range_2006}  that this could be due to a patchy 
distribution of the surfactant 
on the mica surface which creates alternate domains of positive and negative surface charges 
that are able to dynamically self-adjust as the surfaces approach each other. A small scale feature of 
all charged interfaces is that the charge resides at discrete locations on the surface. While it seems 
appropriate to ignore this feature at distances large compared to the scale of this granularity, 
such an assumption may not be valid when considering the interactions at very short range. 
The effect of this granularity in charge distribution has been the 
focus of a number of studies \cite{richmond_electrical_1975,muller_influence_1983,kostoglou_effect_1992,white_increased_2002}. 

In this paper we examine the general problem of the interaction of a pair of infinite parallel 
planes, each with an arbitrary prescribed distribution of surface charge. The 
space between the planes is filled with an electrolyte. 
We derive an expression for the normal 
force acting on either surface by integrating the Maxwell stress over the central plane midway between the two surfaces. We follow Richmond \cite{richmond_electrical_1974} in our use of Fourier transform techniques, but
our approach provides a more direct route to the final result 
than previous investigations based on calculating the free 
energy of the system \cite{richmond_electrical_1974,richmond_electrical_1975,
kostoglou_effect_1992,miklavic_double_1994,white_increased_2002,ben-yaakov_interaction_2013}. 
In particular, we consider the problem of a uniformly charged central section and calculate
the effect of charge overspill on the disjoining pressure. The model of infinite planes
that we adopt here (Figure~\ref{fig:sketch}) allows us to derive analytical results for arbitrary Debye 
length, whereas in earlier work \cite{ghosal_repulsion_2016}
the disjoining pressure between finite blocks was 
found only in the limit of long Debye length. We shall also show how 
our expression for the interaction force
can be used to determine the disjoining pressure 
in other cases, such as periodic or random distributions of charges. 

The paper is organized as follows. In the next section we define our problem and 
show that the case of an arbitrary charge distribution may be
analyzed within the framework of the linearized Poisson-Boltzmann equation
by considering separately the situations
where the charge distributions on the two planes are identical (the symmetric case) and 
where the charge distribution is identical but with opposite sign (the antisymmetric case).
In section~\ref{sec_FourierSpace} the solution for the potential is obtained in terms 
of Fourier transforms with respect to co-ordinates with axes parallel to the planes. In 
section~\ref{sec_NormalForce} we develop an expression for the normal interaction force 
in terms of an integral of the Maxwell stress over the central plane
midway between the charged surfaces. In section~\ref{sec_Applications} we evaluate 
the integral to determine the disjoining pressure for finite charged patches, periodic distributions 
of charge and random charge distribution. In section~\ref{sec_Discussion} we discuss the 
range of validity of our methods, and conclusions are given
in section~\ref{sec_conclusion}.

\section{Formulation}
\label{sec_formulation}

\begin{figure}[t]
   \centering
   \includegraphics[width=0.3\textwidth]{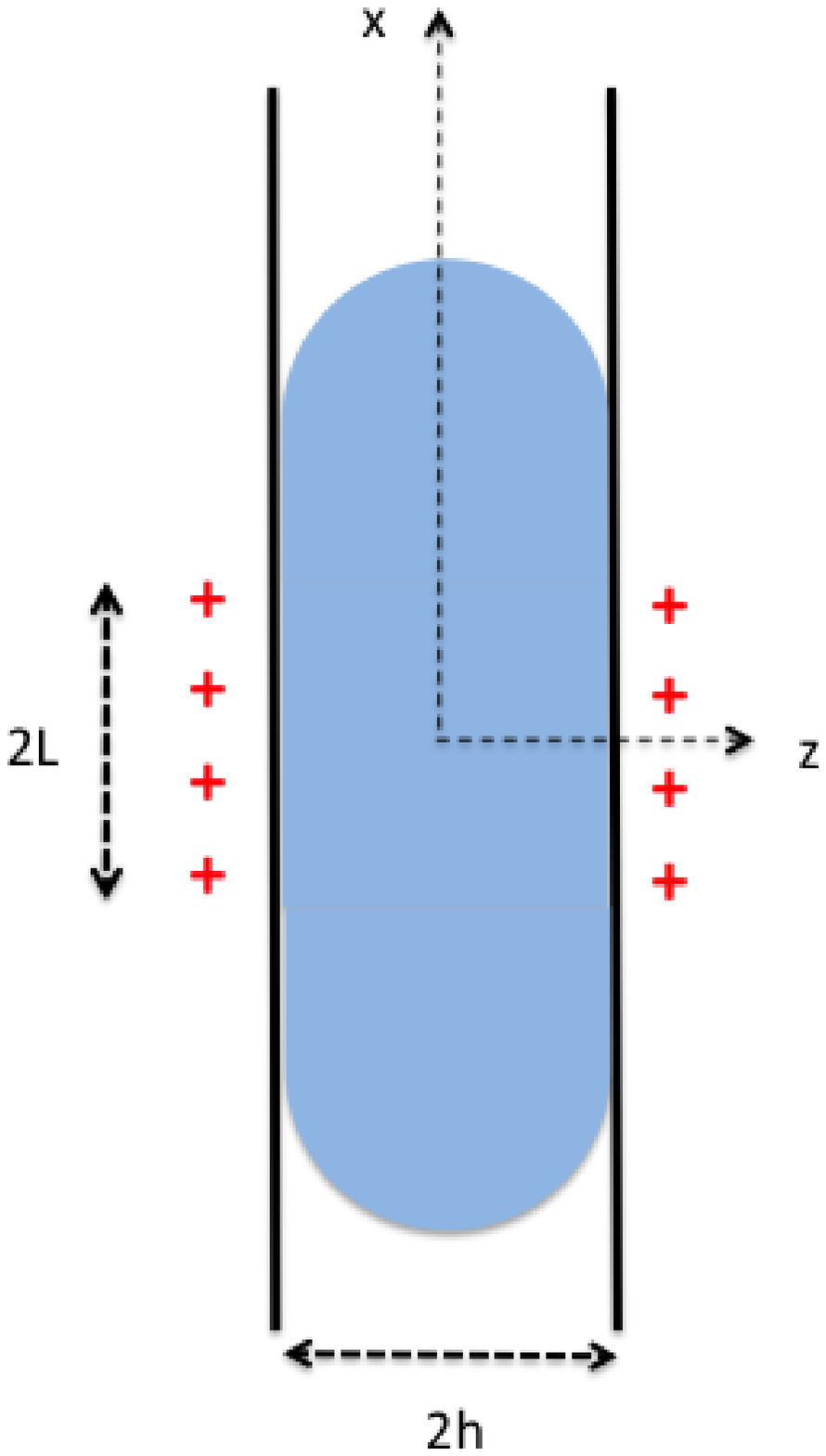} 
   \hspace*{0.05\textwidth}
   \includegraphics[width=0.4\textwidth]{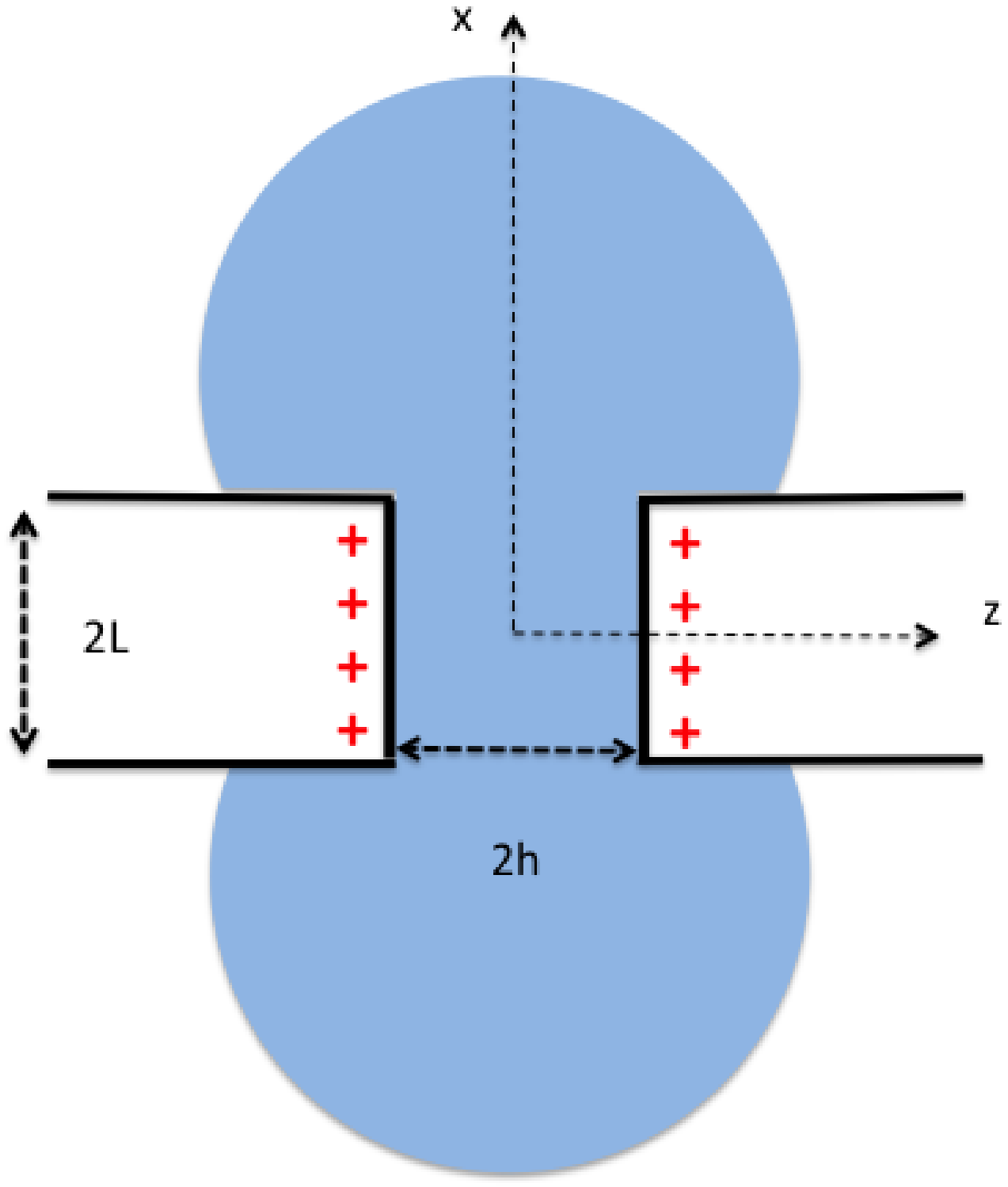} 
   \caption{Sketch showing the geometry of the confined problem (left) and the 
  unconfined problem (right).}
  \label{fig:sketch}
\end{figure}

Consider an electrolyte filled gap of uniform width ($2h$) within a dielectric solid. The electrolyte contains $N$ charged species of valence $z_i$, $i=1,\dots,N$, with $n_i^{(\infty)}$ the equilibrium number density of the $i$ th species far from any charged surfaces.
The charge distributions on the confining walls located
at $z=h$ and $z=-h$ are respectively $\sigma_{+} (x,y)$ and $\sigma_{-}(x,y)$. The geometry 
is as shown in Figure~\ref{fig:sketch} (left panel).
We wish to calculate the component of the force orthogonal to the walls (in the $z$ direction) 
on the plane at $z=h$. Clearly, the normal force on the other 
plane is equal and opposite. 
The problem is treated  in the Debye-H\"{u}ckel limit \cite{russel_colloidal_1989}, in which all potentials are assumed small compared to the thermal
scale $k_BT/e$, where $e$ is the magnitude of the electron charge and $k_BT$
the Boltzmann temperature. The solid 
substrate surrounding the channel is assumed to have permittivity $\epsilon_{s}=0$. This 
approximation is commonly invoked since the relative permittivity of water ($\sim 80$) is much
larger than that of most nonpolar solid substrates ($\sim 2$ -- $4$).
In this section, we assume that charge distributions are 
sufficiently localized so that the functions $\sigma_{+}^{2}$, $\sigma_{-}^{2}$ and 
$\sigma_{+} \sigma_{-}$ are all integrable over the $(x,y)$ plane. This restriction is relaxed 
in section~\ref{sec_Applications} where we consider periodic and random distributions.
In the Debye-H\"{u}ckel limit, the equilibrium potential $\phi$ satisfies
the linearized Poisson-Boltzmann equation
\begin{equation}
\nabla^{2} \phi = \kappa^{2} \phi,
\label{DHeq}
\end{equation}
where $\kappa^{-1}$ is the Debye length.

We define the symmetric case as that in which both 
planes have the charge distribution $\sigma_{S} (x,y)$. In the antisymmetric case,
the charge density is $\sigma_{A} (x,y)$ on $z=h$ and 
$- \sigma_{A} (x,y)$ on $z=-h$.
Thus, the potentials $\phi=\phi_S$ and $\phi=\phi_A$ for the symmetric and antisymmetric 
problems respectively, satisfy (\ref{DHeq}) and the boundary conditions 
\begin{equation} 
\partial_z \phi_S  (x,y,\pm h) = \pm \sigma_{S} (x,y) / \epsilon,
\qquad \partial_z \phi_A (x,y,\pm h) = \sigma_{A} (x,y) / \epsilon,
\label{DHeqBC}
\end{equation}
where $\epsilon$ is the electrolyte permittivity.
For arbitrary charge distributions $\sigma_{+} (x,y)$ on $z=h$ 
and $\sigma_{-}(x,y)$ on $z=-h$ we construct the corresponding symmetric and antisymmetric
charge distributions
\begin{equation}
\sigma_{S} = \frac{1}{2} ( \sigma_{+} + \sigma_{-} ), \qquad
\sigma_{A} =\frac{1}{2} ( \sigma_{+} - \sigma_{-} ). \label{sigmaSA}
\end{equation}
Then, the equilibrium potential is $\phi (x,y,z) =  \phi_{S} (x,y,z) + \phi_{A} (x,y,z)$. 
Indeed, $\phi$ clearly satisfies (\ref{DHeq}) since $\phi_S$ and $\phi_A$
do, and also satisfies the boundary conditions: 
\begin{equation}
 \epsilon \partial_z \phi (x,y,\pm h) =
 \pm \sigma_S + \sigma_A = 
  \begin{cases} 
   \sigma_{+} & \text{if } z = +h,\\
   - \sigma_{-}    & \text{if } z = -h.
  \end{cases}
\end{equation}

\section{Solutions in Fourier space}
\label{sec_FourierSpace}
We use a caret $\; \widehat{\quad} \; $ to indicate Fourier transforms with respect to variables $x$ and $y$.
Thus,
\begin{align} 
\hat{\phi} (k_x,k_y,z) &= 
 \frac{1}{(2 \pi)^{2}} \int  \phi (x,y,z)  \exp[  -i ( k_x x + k_y y) ]  \; dx dy, \\
 \phi (x,y,z) &=  \int  \hat{\phi} (k_x,k_y,z)  \exp[ i ( k_x x + k_y y) ]  \; dk_x dk_y. \label{FTFI}
\end{align} 
Equations (\ref{DHeq}) and (\ref{DHeqBC}) imply that $\hat{\phi}$ satisfies  
\begin{equation} 
\partial_{zz} \hat{\phi} = (k^2 + \kappa^2)  \hat{\phi} \qquad \text{with} \qquad
\epsilon \partial_z \hat{\phi} (k_x,k_y,\pm h) =
  \begin{cases} 
   \pm \hat{\sigma}_{S} \\
   \hat{\sigma}_{A} 
  \end{cases}
\label{BVprob}
\end{equation} 
where $k^2 = k_x^2 + k_y^2$. The two cases in (\ref{BVprob}) correspond to the symmetric 
and anti-symmetric problems respectively.
The solution to (\ref{BVprob}) is readily obtained:
\begin{equation}
\hat{\phi} (k_x,k_y, z) =
  \begin{cases} 
   &\hat{\sigma}_{S} \cosh ( K z) / [ \epsilon K \sinh (K h) ]\\
   &\hat{\sigma}_{A} \sinh ( K z) / [ \epsilon K \cosh (K h) ]
  \end{cases}
\label{BVprobSoln}
\end{equation}
where $K = ( \kappa^{2} + k^{2} )^{1/2}$. 

\section{The normal force} 
\label{sec_NormalForce}
The $z$-component of the force is given as the integral of the $zz$ component
$\Sigma_{zz}$ of the total stress over the the central plane $S$ ($z=0$)
between the plates \cite{russel_colloidal_1989}
\begin{equation}
F_z = - \int_S \Sigma_{zz} \; dx dy, \qquad 
\Sigma_{zz} = - \left. p_{\rm{os}}  \right|_{z=0}
- \frac{\epsilon}{2} |\bm{\nabla}_{h} \phi_{0}|^{2} 
+ \frac{\epsilon}{2} | \phi_{0}^{\prime} |^{2},
\label{Sigmazz1}
\end{equation} 
where $\phi_{0} (x,y) = \phi(x,y,0)$, $\phi^{\prime}_{0} (x,y) = (\partial_{z} \phi)_{z=0}$,
$\bm{\nabla}_{h} = \hat{\bm{i}} \partial_x + \hat{\bm{j}} \partial_y$, and 
\begin{equation}
p_{\rm{os}} = k_{B} T \sum_{i=1}^{N} n_{i}  
= k_{B} T \sum_{i=1}^{N} n_{i}^{(\infty)} \exp(-ez_{i} \phi / k_{B} T) 
\end{equation}
represents the osmotic pressure. 
We have neglected the uniform background pressure  
that gives no contribution to the force.
Equation (\ref{Sigmazz1}) is quite general and is applicable even when the potential $\phi$ 
is not small
compared to the thermal scale, $k_{B} T / e \approx 25$ mV at room temperature.
In the Debye-H\"{u}ckel limit $\phi\ll k_BT/e$ we have the approximate form
\begin{equation}
p_{\rm{os}} = k_{B} T \sum_{i=1}^{N}  n_{i}^{(\infty)} \exp(-ez_{i} \phi / k_{B} T) 
\approx k_{B} T \sum_{i=1}^{N} n_{i}^{(\infty)} + \frac{1}{2} \epsilon \kappa^{2} \phi^{2}.
\label{posDH}
\end{equation}
In the case of a symmetric two component electrolyte $z_{1}=-z_{2}=z$,
$n_{1}^{(\infty)}=n_{2}^{(\infty)}=n_{\infty}$, and thus, 
$p_{\rm{os}} = 2 n_{\infty} k_{B} T \cosh ( e z \phi / k_{B} T )$.
In this paper we discuss the case of low potentials where (\ref{posDH}) is applicable.

\subsection{The symmetric problem} 
\label{subsec_Symmetric}
In the symmetric problem, the last term in the expression for $\Sigma_{zz}$ in 
(\ref{Sigmazz1}) vanishes, and hence,
\begin{equation}
F_z = \frac{\epsilon}{2} \int \left(  \kappa^{2} \phi_{0}^{2} + |\bm{\nabla}_{h} \phi_{0}|^{2} \right)  \; dx dy 
= 2 \pi^2 \epsilon \int K^{2} |\hat{\phi}_{0}|^{2}  \; dk_x dk_y,
\label{Fsymm}
\end{equation}
where the final form follows upon the application of Parseval's identity for Fourier transforms. 
On substituting the solution (\ref{BVprobSoln}) to the symmetric problem, we have
\begin{equation}
F_z = \frac{2 \pi^2}{\epsilon} \int |\hat{\sigma}_{S}|^{2} \cosech^{2} (Kh)  \; dk_x dk_y.
\label{Fzsym}
\end{equation}
\subsection{The antisymmetric problem} 
\label{subsec_ASymmetric}
For the antisymmetric problem, $\phi_{0} (x,y) = 0$ and $\hat{\phi}_{0}^{\prime} = \hat{\sigma}_{A} \sech (Kh) / \epsilon$.
Therefore, 
\begin{equation}
F_{z} = - 2 \pi^2 \epsilon  \int | \hat{\phi}_{0}^{\prime} |^{2} \; dk_x dk_y 
 =  - \frac{2 \pi^2}{ \epsilon } \int | \hat{\sigma}_{A} |^{2} \sech^{2} (Kh) \; dk_x dk_y. 
\label{Fzasym}
\end{equation}
The force is negative, indicating an attractive interaction.
\subsection{The general problem}
\label{subsec_GenProb}
If the charge distributions do not exhibit any special symmetry about the mid plane then in 
place of (\ref{Fsymm}) we have
\begin{equation}
F_z = 2 \pi^2  \epsilon \int \left( K^{2} |\hat{\phi}_{0}|^{2}  - | \hat{\phi}_{0}^{\prime} |^{2} \right) \; dk_x dk_y,
\label{Fgen}
\end{equation} 
since the last $\hat{\phi}_{0}^{\prime}$ term no longer vanishes. On substituting $\phi =  \phi_{S} + \phi_{A}$
in addition to the terms proportional to $| \hat{\phi}_{S} |^{2} $ and $ | \hat{\phi}_{A} |^{2} $ we get a cross term proportional 
to $\hat{\phi}_{S} \hat{\phi}_{A}^{*} + \hat{\phi}_{S}^{*} \hat{\phi}_{A}$ from the first of the two terms in the 
integrand of (\ref{Fgen}), and a term proportional to 
$\hat{\phi}_{S}^{\prime} ( \hat{\phi}_{A}^{\prime} )^{*} + ( \hat{\phi}_{S}^{\prime})^{*} \hat{\phi}_{A}^{\prime}$
from the second term. Clearly, since $\hat{\phi}_{S}^{\prime}$ and $\hat{\phi}_{A}$ both vanish on the 
midplane, the cross terms have a zero contribution to the force. Therefore, the total force may be 
found simply by summing the right hand sides of (\ref{Fzsym}) and (\ref{Fzasym}). Thus, 
we have the following general formula for the force normal to the planes: 
\begin{equation}
\frac{F_z}{2 \pi^2 / \epsilon} = \int \left\{ \frac{|\hat{\sigma}_{S}|^{2}}{\sinh^{2} (Kh)} 
- \frac{| \hat{\sigma}_{A} |^{2}}{\cosh^{2} (Kh)} \right\} dk_x dk_y.
\label{FzGen}
\end{equation}
This is equivalent to the expression for the free energy 
of a pair of parallel 
planes with arbitrary charge distribution found by 
Ben-Yaakov {\it et al.} \cite{ben-yaakov_interaction_2013}.

\subsection{One dimensional distributions}
If $\sigma(x,y)$ is independent of $y$, the charge distributions are no longer square integrable.  
Therefore, (\ref{FzGen}) requires careful interpretion.
In order to do this, we write the charge distribution on either plane as 
\begin{equation}
\sigma_{\pm} (x,y) = \bar{\sigma}_{\pm} (x) \Delta (y) 
\label{2Dto1D}
\end{equation} 
where $\Delta (y)=1$ if $|y| < L_{y}/2$ and $\Delta(y)=0$ otherwise,
$L_y$ being a large but finite length. The 
Fourier transform of (\ref{2Dto1D}) is 
\begin{equation} 
\hat{\sigma}_{\pm} (k_x,k_y) = \hat{\bar{\sigma}}_{\pm} (k_x) \frac{\sin (k_y L_y / 2)}{\pi k_y},
\end{equation} 
where the caret on $\bar{\sigma}_{\pm}$ now indicates the one dimensional Fourier transform 
with respect to $k_x$. Thus, the Fourier transforms of the symmetric and antisymmetric charge densities 
$\sigma_S$ and $\sigma_A$ are also the product of the corresponding 1D transform and the function 
$\sin (k_y L_y / 2) / (\pi k_y )$.  Therefore, in the integration with respect to $k_y$, most of the 
contribution arises from a zone near the origin of width $\sim \pi / L_y$. Since $L_{y}$ is 
large, $K = (\kappa^{2} + k_x^{2} +  k_y^{2})^{1/2} \sim (\kappa^{2} + k_x^{2})^{1/2} \equiv \bar{K}$
and (\ref{FzGen}) may be expressed as a product of integrals
\begin{equation}
 \frac{F_z}{2 \pi^2 / \epsilon}  = \int_{-\infty}^{+\infty}  
\frac{\sin^{2}(k_y L_y / 2)}{\pi^{2} k_y^{2}} \; dk_y 
  \int_{-\infty}^{+\infty} \left\{  \frac{|\hat{\bar{\sigma}}_{S}|^{2}}{\sinh^{2} (\bar{K}h)}
 - \frac{| \hat{\bar{\sigma}}_{A} |^{2}}{\cosh^{2} (\bar{K} h)}  \right\}  \; dk_x.
\label{product_of_integrals}
\end{equation}
The first integral on the right-hand side of (\ref{product_of_integrals}) evaluates to $L_{y} / (2 \pi )$, and thus, 
if we define the force per unit span, $F = F_z / L_y$, then
\begin{equation}
\frac{F}{\pi / \epsilon} = 
 \int_{-\infty}^{+\infty}  \left\{   \frac{|\hat{\sigma}_{S}|^{2}}{\sinh^{2} (Kh)}  
 - \frac{| \hat{\sigma}_{A} |^{2}}{\cosh^{2} (Kh)} \right\}  \; dk,
 \label{genForce1D}
\end{equation}
where we have dropped the bar over the $\sigma$, a one dimensional Fourier transform with 
respect to the $x$-dimension being understood, and $k_x$, $\bar{K}$ 
are replaced by $k$ and $K$ respectively. This is what we would have obtained had we simply 
started by taking a
one-dimensional Fourier transform with respect to $x$ in (\ref{DHeq}).

\section{Applications} 
\label{sec_Applications}
We now consider some applications of (\ref{FzGen}) and (\ref{genForce1D}) 
to special situations where the integral in Fourier space can be evaluated analytically.
\subsection{Uniformly charged section of finite length}
\label{subsec_1D}
When two surfaces with like charge interact across a gap of width $2h$, the interaction force decreases 
exponentially \cite{russel_colloidal_1989}   with $h$
as long as $\kappa h \gg 1$. In the opposite 
limit of $\kappa h \ll 1$ the Debye layers on the two planes are strongly overlapped and the exponential dependence gives way to a power law $\Pi \sim h^{-2}$ 
at small separations \cite{philipse_algebraic_2013}, since the potential and ionic number densities are approximately uniform over the gap width and Donnan equilibrium holds \cite{philipse_donnan_2011}.
If the charged section of the planes is of finite size ($\sim L$), the charge cloud between the planes tends to spill out (`overspill') beyond
this region  \cite{ghosal_repulsion_2016}. This leads
to an edge correction to the interaction force which can be non-negligible
unless $\kappa L\gg 1$.

We consider two problems. In the first problem the planes are unbounded, with uniform separation $2h$. 
The charge density on either plane is
$\sigma(x) = \sigma_{0}$ if $|x| < L$ and zero otherwise.
Since the planes have infinite extent, the charge cloud of counter ions is confined in 
the gap $-h<z<h$, though some of it overspills the charged section $|x| < L$. Thus, there is a loss of confinement of the Debye layer in the direction parallel to the planes but not in the normal direction. For brevity, we will call this 
the ``confined problem''. 
In the unconfined problem, the planes are again uniformly
charged in the region
$|x|<L$, where the gap width is $2h$. However, the planes are
of finite extent, and in $|x|>L$ the region $-\infty<z<\infty$ is
occupied by electrolyte. The charge cloud of counter ions is confined in
$-h<z<h$ in $|x|<L$, but is unconfined where it spills out into $|x|>L$.
The two cases are shown schematically in Figure~\ref{fig:sketch}.

The unconfined geometry is the more realistic of the two, and has been studied
\cite{ghosal_repulsion_2016} using asymptotic methods that
are appropriate only in the limit $\kappa h \ll 1$.
The confined problem, though perhaps less realistic, can be solved exactly
within the Debye-H\"{u}ckel limit for any value of $\kappa h$, and
will be studied in detail below.
These two problems can be considered as examples of a more general problem
in which both the surface charge density and separation between the planes
are functions of $x$.

\subsubsection{Like charge} \label{subsubsec_uniform_sec_like}
We consider the confined problem in which $\sigma$ is uniform 
over a central section and zero elsewhere,
with identical distribution on the two planes. 
The Fourier transform of the charge density is 
\begin{equation}
\hat{\sigma} (k) = \frac{1}{2 \pi} \int_{-\infty}^{+\infty} \sigma (x) \exp (- i k x) \; dx 
= \frac{1}{2 \pi} \int_{-L}^{+L} \sigma_0 \exp (- i k  x) \; dx
= \frac{\sigma_0}{\pi k} \sin( k L).
\end{equation}
Since the charge is symmetric with respect to the midplane in this problem, $\sigma_S = \sigma$ 
and $\sigma_A = 0$. Thus, from (\ref{genForce1D}), the force on the plane $z=h$ is 
\begin{equation}
F = \frac{\sigma_{0}^{2}}{\pi \epsilon} \int_{-\infty}^{+\infty}  \frac{\sin^{2} (kL)}{k^{2} \sinh^{2} (Kh)}  \; dk.
\end{equation}
We make a change of variables to $\eta = kL$ in this integral. The disjoining pressure $\Pi = F / (2L)$ is therefore 
\begin{equation} 
\frac{\Pi}{ \sigma_{0}^{2} / (2 \epsilon) } = 
\frac{1}{\pi} \int_{-\infty}^{+\infty}  
 \frac{\sin^{2} \eta }{\eta^2
 \sinh^{2}  \left( \kappa h \sqrt{1  + \eta^{2} / \kappa^2 L^2 } \right) } \; d \eta.
 \label{eq4Pi}
 \end{equation} 
 The integral on the right may be evaluated 
 numerically for specific pairs of values of $\kappa h$ and $\kappa L$. It is however
 instructive to first study some special limits. 
Taking the limit $\kappa L \rightarrow \infty$ in (\ref{eq4Pi}) we have 
 \begin{equation} 
 \frac{\Pi}{ \sigma_{0}^{2} / (2 \epsilon) }  =
\frac{ \cosech^{2}  ( \kappa h ) }{\pi} \int_{-\infty}^{+\infty} 
 \frac{\sin^{2} \eta }{\eta^2} \; d \eta 
 =  \cosech^{2}  ( \kappa h ) = \frac{\Pi_{\infty}}{ \sigma_{0}^{2} / (2 \epsilon) },
 \label{PiSymInf}
 \end{equation} 
which is the classical result  for two uniformly 
charged infinite planes. 
Thus, when $\kappa h$ is large, $\Pi_{\infty}  \sim   \exp ( - 2  \kappa h )$
corresponding to the weak overlap approximation \cite{russel_colloidal_1989},
and when $\kappa h$ is small, $\Pi_{\infty} \sim (\kappa h)^{-2}$ 
\cite{philipse_algebraic_2013}.

\begin{figure}[t]
    \centering
    \includegraphics[width=0.48\textwidth]{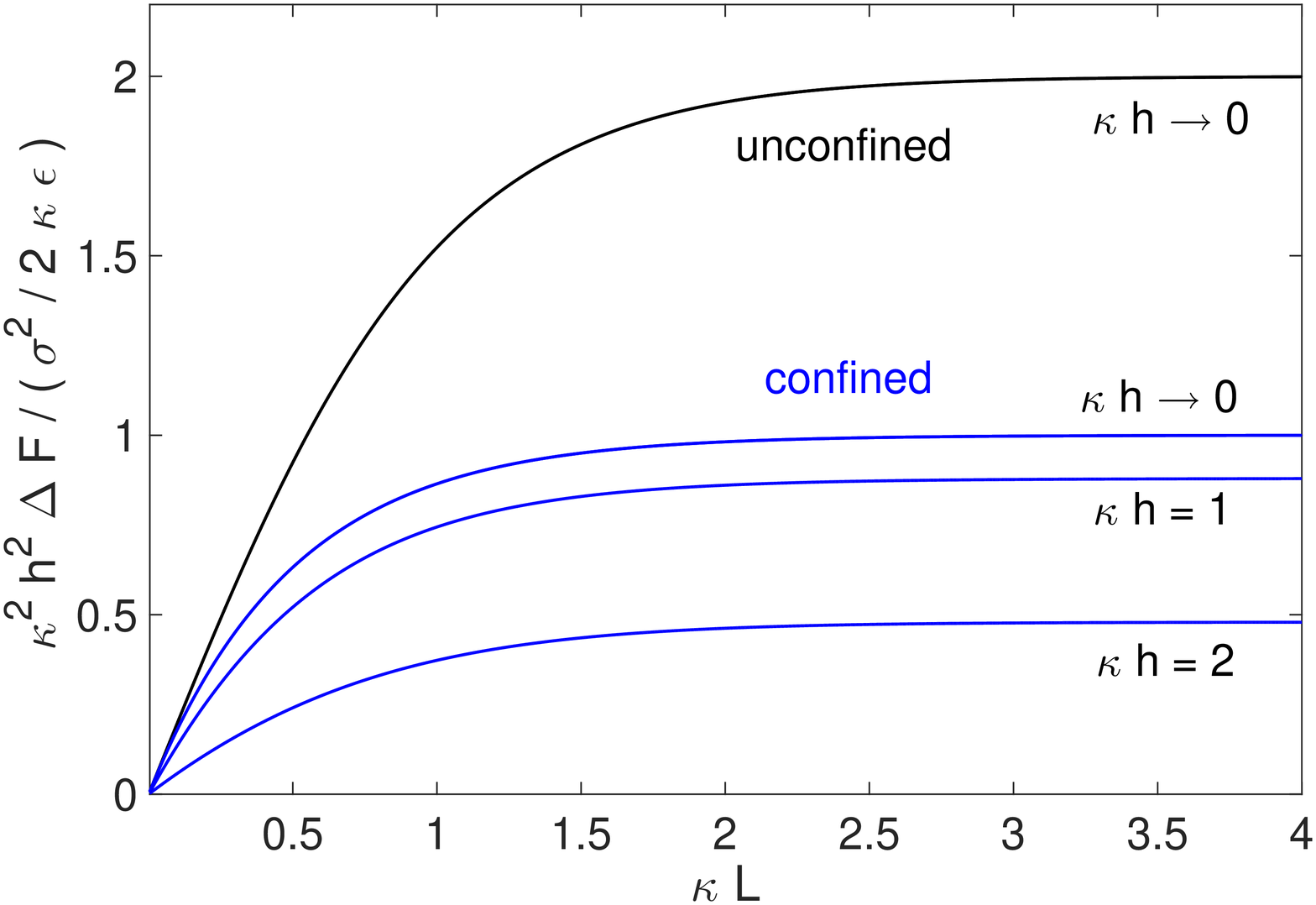} 
   \includegraphics[width=0.48\textwidth]{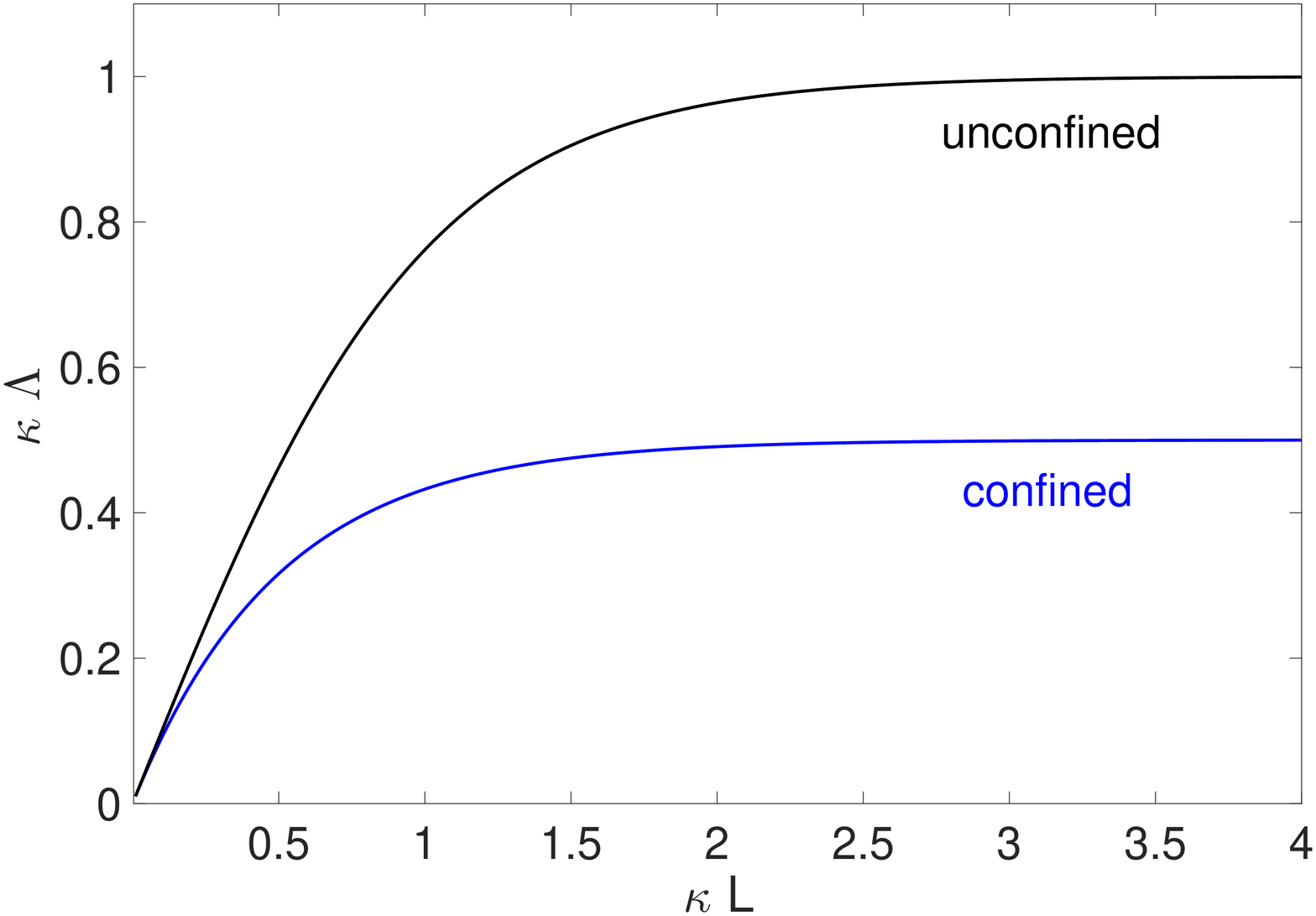}
    \caption{{\it Left panel:} The normalized force deficit as a function of $\kappa L$ for: 
   (i) the unconfined problem (\ref{DeltaFA}) in the limit $\kappa h \rightarrow 0$; 
(ii) 
   the confined problem (\ref{DeltaFB}) in the limit $\kappa h \rightarrow 0$; 
(iii)
  the confined problem  (\ref{eq4Pi}) with $\kappa h = 1$ and $2$.
  {\it Right panel:} The normalized lost length $\kappa \Lambda$ as a function of the normalized 
    plate length $\kappa L$ in the limit $\kappa h \rightarrow 0$ for: (i) the unconfined problem 
    (\ref{unconfined_lost_length});
    (ii) the confined problem, (\ref{PB_lostlength_maintext}).
     }
    \label{fig:kapLambda}.
 \end{figure}

We now consider the limit $\kappa h \rightarrow 0$ at fixed $\kappa L$.
One may then make the approximation $\sinh x \sim x$ in (\ref{eq4Pi}), so that 
\begin{equation} 
\kappa^2 h^2 \frac{\Pi}{ \sigma_{0}^{2} / (2 \epsilon) } \sim 
\frac{1}{\pi \kappa L} \int_{-\infty}^{+\infty}  
 \frac{\sin^{2} (\kappa L \xi)}{\xi^2 (1 + \xi^{2})} \; d \xi.
 \label{eq4PiA}
 \end{equation} 
 The  integral may be evaluated exactly (see Appendix), 
 and thus, we have 
 \begin{equation} 
\frac{\Pi}{ \sigma_{0}^{2} / (2 \epsilon) } \sim 
\frac{1}{\kappa^{2} h^{2}} \left[ 1 - \frac{1}{2 \kappa L} \left\{ 
1 - \exp(- 2 \kappa L ) \right\} \right],\quad \kappa h\ll 1.
 \label{eq4Pismallh}
 \end{equation}
 Let us define the force deficit due to edge effects as $\Delta F = 2L (\Pi_{\infty} - \Pi)$. Then (\ref{eq4Pismallh}) 
 may be expressed as 
 \begin{equation}
 \kappa^{2} h^{2} \frac{\Delta F}{\sigma_{0}^{2}/(2 \epsilon \kappa)}  \sim  1 - \exp(-2 \kappa L) ,\quad \kappa h\ll 1.
 \label{DeltaFB}
 \end{equation} 
 When $\kappa L$ is large, $\Delta F \sim \sigma_{0}^{2}/(2 \epsilon \kappa^{3} h^{2})$,
 independent of $L$. The equivalent of (\ref{DeltaFB}) for the unconfined
problem is \cite{ghosal_repulsion_2016}
 \begin{equation} 
 \kappa^{2} h^{2} \frac{\Delta F}{\sigma_{0}^{2}/(2 \epsilon \kappa)}  \sim  2 \tanh ( \kappa L),\quad \kappa h\ll 1.
 \label{DeltaFA}
 \end{equation} 
 Thus, in the limit of large $\kappa L$, the force deficit in the unconfined problem is a factor of 2 
 larger than that for the confined problem. This increase in the force deficit is to be expected, since
there is more scope for ions to spill out of the charged central region
in the unconfined geometry than in the confined geometry (Figure~\ref{fig:sketch}). The force deficits in the confined and unconfined 
cases are shown in Figure~\ref{fig:kapLambda}.
The reduction in the disjoining force is due to the spillage of the Debye charge cloud from the 
gap between the planes. The amount of this 
spillage can be characterized by the ``lost length'' $\Lambda$,
defined \cite{ghosal_repulsion_2016} 
as the 
distance along the plane
 that would normally confine an amount of Debye 
layer charge equal to the charge $Q$ spilled out beyond one of the edges.
For the confined problem, in the limit $\kappa h\rightarrow 0$, 
\begin{equation}
\Lambda=-\frac{Q}{2\sigma}=
\frac{1}{\kappa\left\lbrack 1+\coth(\kappa L)\right\rbrack}
\label{PB_lostlength_maintext}
\end{equation}
(see Appendix).
From (\ref{PB_lostlength_maintext}) we see that
$\Lambda \sim L$ or $\kappa^{-1}/2$ respectively in the limits of short 
($\kappa L \ll 1$) and long ($\kappa L \gg 1$) plate lengths.
This must be expected, since when the length of the charged section  
approaches zero, all of the Debye cloud is 
spilled outside its confines whereas when $L\gg \kappa^{-1}$ only the charge confined within 
a distance of the order of a Debye length of the edge is lost. In the latter case, the lost charge is a small fraction of 
the total charge in the counter-ion cloud and the force deficit is also a small correction,
independent of $L$.
These results can be compared to the corresponding result \cite{ghosal_repulsion_2016}
for the unconfined problem, for which,
in the limit $\kappa h \rightarrow 0$,
\begin{equation}
\Lambda = \kappa^{-1} \tanh (\kappa L).
\label{unconfined_lost_length}
\end{equation}
The lost lengths for the unconfined and confined problems
in the limit $\kappa h \rightarrow 0$ 
are compared in Figure~\ref{fig:kapLambda}. The lost length is smaller 
in the confined problem which allows less scope for ions to spill out of the
gap between the charged 
sections. 

\subsubsection{Unlike charge}
\label{subsubsec_uniform_sec_unlike}
Let us revisit the problem of a uniformly charged finite section, except that the charges 
on the two planes are now assumed to be equal and opposite. Thus,
$\sigma (x) = \pm \sigma_0$ (corresponding to the planes $z=\pm h$)
if $|x| < L$ and zero otherwise.
 In this case, $\sigma_S = 0$ and $\sigma_A (x) = |\sigma (x)|$. 
Equation (\ref{genForce1D}) then implies 
\begin{align}
F &= - \frac{\pi}{\epsilon} \int_{-\infty}^{+\infty}  \frac{| \hat{\sigma} |^{2}}{\cosh^{2} (Kh)}  \; dk 
= - \frac{\sigma_{0}^{2}}{\epsilon \pi} \int_{-\infty}^{+\infty}  
\frac{\sin^{2} (kL)}{k^{2} \cosh^{2} (Kh)} \; dk \nonumber\\
&= - \frac{\sigma_{0}^{2} L}{\epsilon \pi}  \int_{-\infty}^{+\infty}  
\frac{\sin^{2} \eta \quad }{\eta^{2} \cosh^{2} \left( \kappa h \sqrt{ 1 + \eta^{2} / \kappa^2 L^2 } \right) }
\; d \eta.
\end{align}
Thus, the disjoining pressure, $\Pi = F / (2 L)$ in this case is
\begin{equation} 
 \frac{\Pi}{ \sigma_{0}^{2} / (2 \epsilon) } = 
 - \frac{1}{\pi} \int_{-\infty}^{+\infty}  
 \frac{\sin^{2} \eta \quad  }{\eta^2
 \cosh^{2}  \left( \kappa h \sqrt{1  + \eta^{2} / \kappa^2 L^2 } \right) } \; d \eta. 
\label{eq4Pi_unlike}
\end{equation} 
The disjoining pressure is negative (attractive interaction), and can be evaluated by numerical integration 
for any pair of values of $\kappa h$ and $\kappa L$. 
However, we again first study some special limits.
Taking the limit $\kappa L \rightarrow \infty$ in (\ref{eq4Pi_unlike}) 
 we have 
 \begin{equation} 
 - \frac{\Pi}{ \sigma_{0}^{2} / (2 \epsilon) } = 
\frac{ \sech^{2}  ( \kappa h ) }{\pi} \int_{-\infty}^{+\infty} 
 \frac{\sin^{2} \eta }{\eta^2} \; d \eta = \sech^{2}  ( \kappa h ) = - \frac{\Pi_{\infty}}{ \sigma_{0}^{2} / (2 \epsilon) }
 \label{eq4PiLinf1}
 \end{equation} 
 for  two uniformly charged infinite planes with unlike charge. In this case,
 $\Pi_{\infty} \sim \exp(- 2 \kappa h)$ when $\kappa h$ is large
 but in the limit of vanishing separation, $\kappa h \rightarrow 0$, we have a finite disjoining pressure.
 \subsection{Two dimensional periodic distributions}
 Let us now suppose that we have a charge distribution $\sigma (x,y)$ that is a periodic function of $x$
 and $y$. With no loss of generality we may assume that the charge distribution is generated by 
 specifying it in a finite domain 
 $D_{0} = [-L_{x}/2,L_{x}/2] \times  [-L_{y}/2,L_{y}/2]$ which is then repeated $(2N+1)$ times in the $x$ and $y$ 
 directions. The value of $\sigma$ is taken as zero outside the $(2N+1)L_{x} \times (2N+1) L_{y}$ sized 
 rectangular region. Thus, $\sigma (x,y)$ is of compact support and square integrable. We will of course 
 pass to the limit $N \rightarrow \infty$ in the final answer; we will call $D_{0}$ the primitive cell.

 We temporarily suppress the subscript `S' or `A' and simply use $\sigma$ for the charge distribution. 
 On account of the periodicity we have 
 \begin{align} 
 \hat{\sigma} (k_x,k_y) &= \tilde{\sigma} (k_x,k_y) 
    \frac{L_x L_y}{ (2  \pi)^2}  \sum_{m=-N}^{+N}  \sum_{n=-N}^{+N}
 \exp[ - i (m k_x L_x + n k_y L_y ) ]  \nonumber\\
 &= \tilde{\sigma} (k_x,k_y)  \frac{L_x L_y}{ (2 \pi)^2} g \left( \frac{k_x L_x}{2} \right) g \left( \frac{k_y L_y}{2} \right).
 \end{align} 
 Here $\tilde{\sigma}$ is defined as 
\begin{equation} 
 \tilde{\sigma} (k_x,k_y) = 
\frac{1}{L_x L_y} \int_{D_{0}}  \sigma(x,y)  \exp[ -i ( k_x x + k_y y) ] \;  dx dy
\end{equation}
and the function 
\begin{equation} 
g (\xi) = \sum_{m=-N}^{+N} \exp ( -2 i m \xi ) 
= \frac{\sin \{ (2N+1) \xi \}}{\sin \xi}.
\end{equation} 
The function $g (\xi)$ has the following properties when $N$ is large:
(i) when $\xi \rightarrow  \xi_k = k \pi$ ($k=0,\pm 1, \pm 2, \cdots $), $g(\xi) \rightarrow 2N+1$,
(ii) for all other values $g (\xi)$ oscillates rapidly about zero with an amplitude of order unity,
(iii) the integral of the square of $g (\xi)$ in the neighborhood of $\xi_k$ is $(2N + 1) \pi$. 
Indeed,
\begin{equation}
 \int_{\xi_k - 0}^{\xi_k + 0} g^{2} (\xi) \, d \xi  
\sim \int_{-\infty}^{+\infty} \frac{\sin^{2} \{ (2N+1) (\xi - \xi_k ) \} }{ (\xi - \xi_k)^{2} } \, d \xi
=(2N+1) \int_{-\infty}^{+\infty} \frac{\sin^{2} x}{x^2} \, dx = (2N+1) \pi.
\end{equation}
Thus, in the neighborhood of $\xi = \xi_k$, in the limit $N \rightarrow \infty$.
\begin{equation} 
g^2 ( \xi ) \rightarrow \pi (2N + 1) \delta (\xi - \xi_k ) 
\end{equation} 
where $\delta$ denotes the Dirac delta function. Let us define the reciprocal lattice as a set of wave vectors 
$\bm{\rho} = \hat{\bm{i}} (2 \pi m / L_x) + \hat{\bm{j}} (2 \pi n / L_y)$ 
where $m,n = 0, \pm 1, \pm 2, \pm 3, \cdots$. Then at any neighborhood of a reciprocal lattice vector,
\begin{align}
 g^{2} \left( \frac{k_x L_x}{2} \right)  g^{2} \left( \frac{k_y L_y}{2} \right) 
&= \pi^2 (2N + 1)^2 \delta \left( \frac{k_x L_x}{2} - m \pi \right) \delta \left( \frac{k_y L_y}{2}
 - n \pi \right) \nonumber\\
 &\rightarrow \frac{4 \pi^2 (2N+1)^{2}}{L_x L_y} \delta ( {\bm k} - {\bm \rho}).
\end{align}
Thus, the integral in (\ref{FzGen}) reduces to a sum over the reciprocal lattice and we have 
for the disjoining pressure
\begin{equation}
\Pi = \frac{F_z}{L_x L_y (2N+1)^{2}} 
=  \frac{1}{2 \epsilon} \sum_{{\bm \rho}}  \left[  
\frac{|\tilde{\sigma}_{S}|^{2}}{\sinh^{2} (Kh)}   - \frac{| \tilde{\sigma}_{A} |^{2}}{\cosh^{2} (Kh)}  \right]
\label{DP_per}
\end{equation}
where $K = ( \kappa^2 + \rho^2 )^{1/2}$.

A uniformly charged plate may be considered a periodic charge with the period $L_x, L_y \rightarrow \infty$.
In this case, the reciprocal lattice contains only the single point ${\bm \rho} = 0$ where 
$\tilde{\sigma} (k_x,k_y) =  \sigma_0$. 
Thus, from (\ref{DP_per}),
$\Pi = \Pi_{\text{sym}} = ( \sigma_{0}^{2} / 2 \epsilon) \cosech^{2} ( \kappa h )$
in the symmetric case and 
$\Pi = \Pi_{\text{asym}} = - ( \sigma_{0}^{2} / 2 \epsilon) \sech^{2} ( \kappa h )$
in the anti-symmetric case, as expected. 

\subsubsection{Zebra stripes} 
Let us suppose that we have charged stripes of width $\delta$ parallel to the $y$-axis separated 
by uncharged sections of width $\Delta$. In this case, the primitive cell may be taken as the domain 
$[-L_x /2,L_x /2] \times [-L_y/2,L_y/2]$ 
where $L_x = \delta + \Delta = L$ and $L_y \rightarrow \infty$. 
The interval $- \delta/2 >  x > \delta / 2$ carries a charge density 
$\sigma_0$ and the remainder of the cell has zero charge. The reciprocal 
lattice then consists of the vectors 
${\bm \rho} = \hat{{\bm i}} (2 \pi / L) n$, where $n=0,\pm 1,\pm 2, \cdots$, and,
\begin{equation} 
\tilde{\sigma} ({\bm \rho}) = \frac{\sigma_0}{L} \int_{-\delta/2}^{\delta/2} \exp \left( - 2 n \pi i
\frac{x}{L} \right) \; dx
= \frac{\sigma_{0}}{n \pi} \sin \left( \frac{n \pi \delta}{L} \right).
\end{equation} 
On substituting in (\ref{DP_per}), we have in the symmetric case 
\begin{equation} 
 \frac{\Pi_{\text{sym}}}{\sigma_{0}^{2} / (2 \epsilon) }
= \frac{\delta^2}{L^2} \cosech^{2} (\kappa h) 
+ \sum_{n=1}^{\infty} \frac{2}{n^2 \pi^2} \sin^{2} \left( \frac{n \pi \delta}{L} \right) 
  \cosech^{2} \left( \kappa h \sqrt{1 + \frac{4 \pi^2 n^2}{\kappa^2 L^2}} \right),
\end{equation} 
and in the antisymmetric case 
\begin{equation} 
 - \frac{\Pi_{\text{asym}}}{\sigma_{0}^{2} / (2 \epsilon) } 
=  \frac{\delta^2}{L^2} \sech^{2} (\kappa h)
+ \sum_{n=1}^{\infty} \frac{2}{n^2 \pi^2} \sin^{2} \left( \frac{n \pi \delta}{L} \right)  
 \sech^{2} \left( \kappa h \sqrt{1 + \frac{4 \pi^2 n^2}{\kappa^2 L^2}} \right).
\end{equation} 
Note that if $\Delta \rightarrow 0$, $\delta \rightarrow L$ and in this case we recover the 
result for the disjoining pressure between uniformly charged planes.
\subsubsection{Zebra stripes with alternating charge} 
Instead of having  positively charged stripes separated by uncharged regions, let us now 
suppose that we have alternate bands of positive charge (surface density $\sigma_0$, width 
$\Delta_{+}$) and negative charge (surface density $-\sigma_0$, width $\Delta_{-}$).
The primitive cell once again is of width $L = \Delta_{+} + \Delta_{-}$, of which the 
central region $(- \Delta_{+}/2,\Delta_{+}/2)$ carries a surface charge density of $\sigma_0$ 
with the remainder of the cell carrying $- \sigma_0$. As before, $L_y = \infty$. 
Thus, 
\begin{align}
\tilde{\sigma} ({\bm \rho}) &= \frac{\sigma_0}{L} 
\left(
\int_{-\Delta_{+}/2}^{\Delta_{+}/2}  
- \int_{-L/2}^{- \Delta_{+}/2}
- \int_{\Delta_{+}/2}^{L/2} 
\right) e^{- 2 i n \pi x / L } \; dx   \nonumber\\
&= \frac{\sigma_0}{L}
 \begin{cases} 
   \Delta_{+} - \Delta_{-}  & \text{if } n=0, \\
   (2 L / n \pi)  \sin( n \pi \Delta_{+} / L ) & \text{if } n > 0.
  \end{cases}
\end{align}
On substituting in (\ref{DP_per}), we have in the symmetric case 
\begin{equation}
 \frac{\Pi_{\text{sym}}}{\sigma_{0}^{2} / (2 \epsilon) }
= \left( \frac{\Delta_{+} - \Delta_{-}}{\Delta_{+} + \Delta_{-}} \right)^{2} 
\cosech^{2} (\kappa h) 
+ \sum_{n=1}^{\infty} 
\frac{8 \sin^{2} ( n \pi \Delta_{+} / L )}{n^2 \pi^2
\sinh^{2} \left( \kappa h \sqrt{1 + \frac{4 \pi^2 n^2}{\kappa^2 L^2} }  \right)},
\label{zebra_sym}
\end{equation}
and in the antisymmetric case 
\begin{equation} 
- \frac{\Pi_{\text{asym}}}{\sigma_{0}^{2} / (2 \epsilon) }
= \left( \frac{\Delta_{+} - \Delta_{-}}{\Delta_{+} + \Delta_{-}} \right)^{2} 
\sech^{2} (\kappa h) + \sum_{n=1}^{\infty}
 \frac{ 8 \sin^{2} ( n \pi \Delta_{+} / L ) }{
  n^2 \pi^2 \cosh^{2} \left( \kappa h \sqrt{1 + \frac{4 \pi^2 n^2}{\kappa^2 L^2}} \right)}.
\label{zebra_asym}
\end{equation} 
Note that if either $\Delta_{+}=0$ or $\Delta_{-} = 0$, 
the result for uniformly charged planes is recovered. Further note that if $\Delta_{+} = \Delta_{-}$
then both planes are overall charge neutral. In this case, the first terms in (\ref{zebra_sym}) 
and (\ref{zebra_asym}) vanish. But since $\cosech x > \sech x$, each term in the series 
(\ref{zebra_sym}) exceeds in magnitude the corresponding term in the series 
(\ref{zebra_asym}). Thus, 
\begin{equation} 
| \Pi_{\text{sym}} | > | \Pi_{\text{asym}} |
\end{equation}
in agreement with \cite{ben-yaakov_interaction_2013}.

\subsubsection{Checker board pattern} 
We now consider a ``checker board pattern'' where each square is of edge length $\Delta$ and 
alternate cells have charge density $\pm \sigma_0$. The primitive cell may then be taken as a 
square of edge length $2 \Delta$ divided into four subunits. The first and third quadrants carry 
a charge density $\sigma_0$ whereas the remaining quadrants carry a charge density 
$- \sigma_{0}$. The reciprocal lattice consists of the vectors 
${\bm \rho} =  \hat{{\bm i}} (m \pi / \Delta) + \hat{{\bm j}} ( n \pi / \Delta)$
where $m,n=0,\pm 1, \pm 2, \cdots$. A straightforward calculation shows that 
\begin{equation} 
\tilde{\sigma} ( {\bm \rho} ) = 
\begin{cases} 
   - 4 \sigma_{0} / \pi^2 m n  & \text{if $m$ and $n$ are odd,} \\
   0  & \text{otherwise.}
  \end{cases}
\end{equation} 
On substituting in (\ref{DP_per}), we have in the symmetric case 
\begin{equation}
\frac{\Pi_{\text{sym}}}{\sigma_{0}^{2} / (2 \epsilon) } =  \frac{64}{\pi^4}  
\sum_{\substack{m,n=1 \\ \text{$m$,$n$ odd}}}^{\infty} \frac{1}{m^2 n^2} 
\cosech^{2} \left( \kappa h \sqrt{1 + \frac{\pi^2}{\kappa^2 \Delta^2} (m^2 + n^2) } \right).
\end{equation}
The antisymmetric case corresponds to shifting the planes relative to each other  
in either the $x$ or $y$ directions by the amount $\Delta$. The disjoining pressure 
in this case is negative and we have 
\begin{equation}
 - \frac{\Pi_{\text{asym}}}{\sigma_{0}^{2} / (2 \epsilon) }
= \frac{64}{\pi^{4}} \sum_{\substack{m,n=1 \\ \text{$m$,$n$ odd}}}^{\infty} 
\frac{1}{m^2 n^2} 
\sech^{2} \left( \kappa h \sqrt{1 + \frac{\pi^2}{\kappa^2 \Delta^2} (m^2 + n^2) } \right).
\end{equation}
\subsubsection{Point charges on square lattice} 
\begin{figure}[t]
   \centering
   \includegraphics[width = 0.48\textwidth]{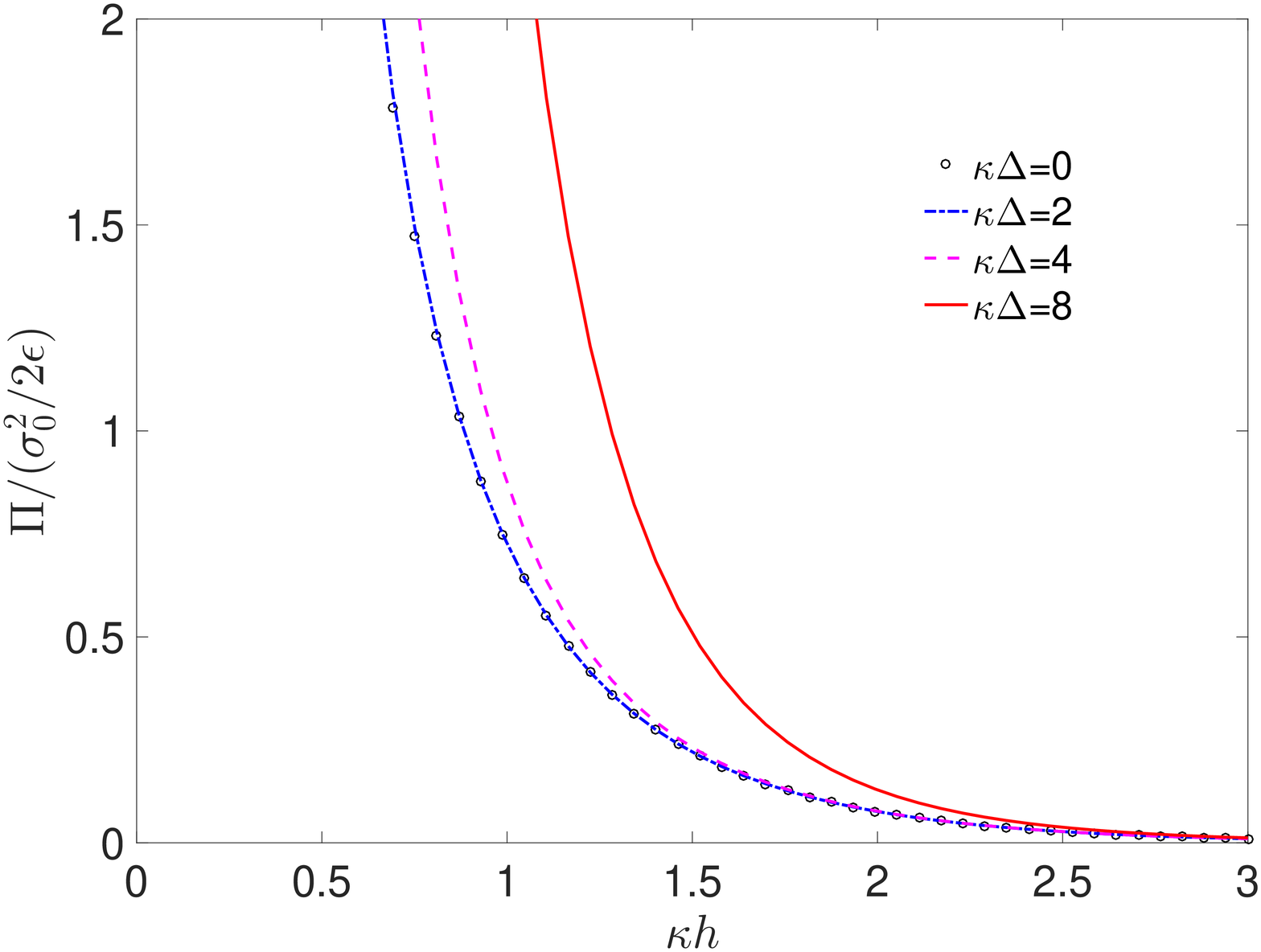} 
   \includegraphics[width = 0.48\textwidth]{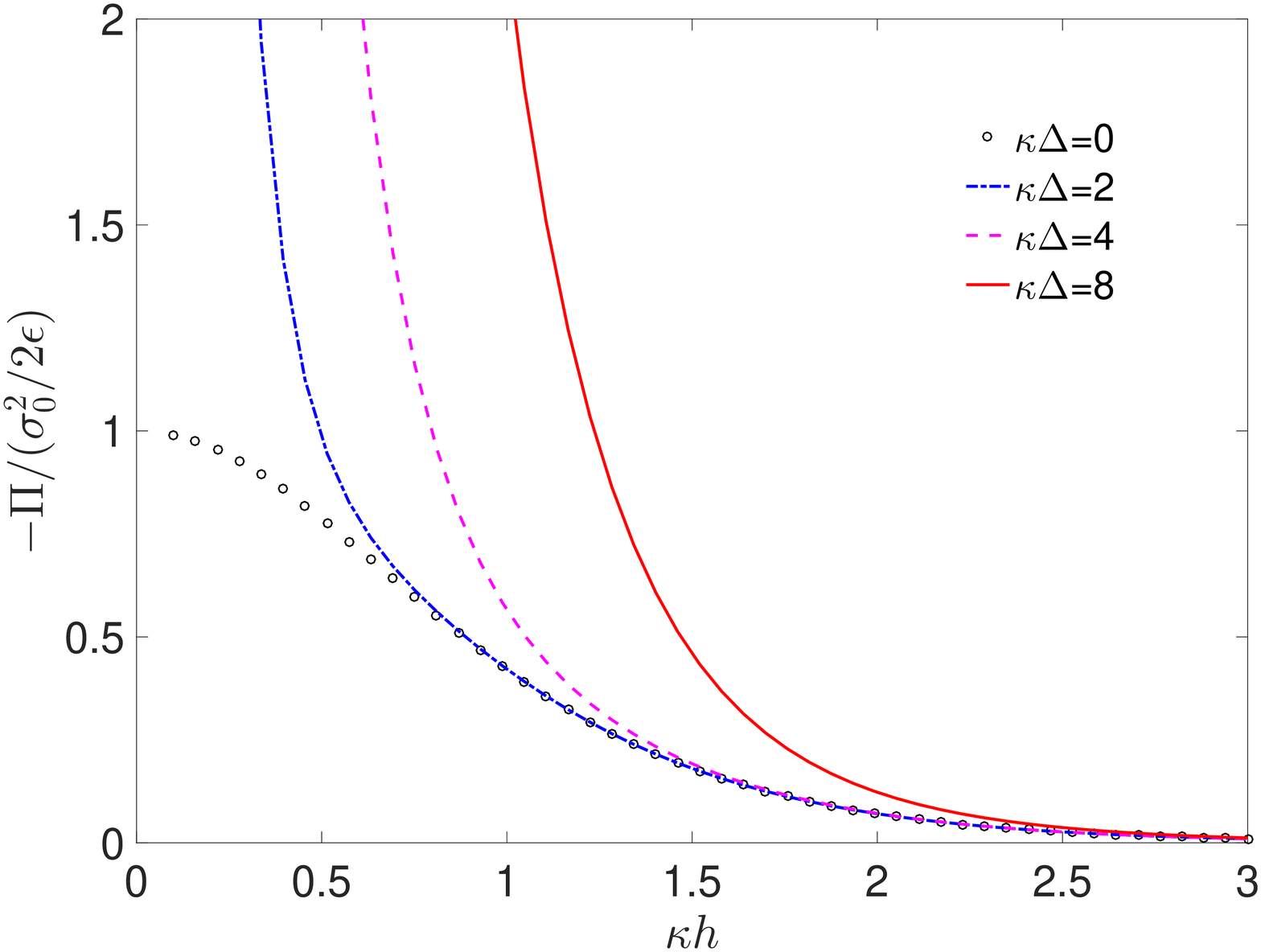} 
   \caption{Normalized disjoining pressure between a pair of flat surfaces with discrete point 
   charges on a square lattice 
   as a function of the dimensionless plate separation ($\kappa h$) for fixed values of the 
   normalized lattice  spacing ($\kappa \Delta$). The symmetric (left) and 
   antisymmetric (right) cases are shown.
  The limit $\kappa \Delta = 0$ corresponds to a uniform distribution with no granularity.}
  \label{fig:wigner}
\end{figure}

Let us consider a two dimensional array of point charges, $Q$, located at the nodes of a 
square lattice of spacing $\Delta$. This could be a representation of the discrete
nature of the charge distribution on surfaces when viewed on atomic scales.
Here the primitive cell may be taken 
as the square $[-\Delta/2,\Delta/2] \times [-\Delta/2,\Delta/2]$ with a charge $Q$ at the 
origin. The reciprocal lattice consists of the vectors 
${\bm \rho} =   \hat{{\bm i}} (2 \pi m / \Delta) + \hat{{\bm j}} (2 \pi n / \Delta)$
where $m,n=0,\pm 1, \pm 2, \cdots$, and 
\begin{equation} 
 \tilde{\sigma} ({\bm \rho}) = \frac{1}{\Delta^2} \int_{D_{0}}  Q \delta ({\bm r})  e^{-i {\bm \rho}
 \cdot {\bm r} } \;  d^{2} {\bm r} = \frac{Q}{\Delta^2}.
\end{equation}
On substituting in (\ref{DP_per}), we have in the symmetric case 
\begin{equation}
\frac{\Pi_{\text{sym}}}{\sigma_{0}^{2} / (2 \epsilon) }
= \cosech^{2} (\kappa h) +  \sum_{m,n=1}^{\infty}
4 \cosech^{2} \left( \kappa h \sqrt{1 + \frac{4 \pi^2}{\kappa^2 \Delta^2} (m^2 + n^2) } \right),
\label{WignerS}
\end{equation}
where $\sigma_0 = Q / \Delta^2$ is the average charge density.
The anti-symmetric case is also interesting as it describes, for example, 
a surface with a discrete charge distribution approaching a conducting 
plane. The corresponding result for the disjoining pressure is 
\begin{equation}
 - \frac{\Pi_{\text{asym}}}{\sigma_{0}^{2} / (2 \epsilon) }
=  \sech^{2} (\kappa h) + \sum_{m,n=1}^{\infty} 
 4 \sech^{2} \left( \kappa h \sqrt{1 + \frac{4 \pi^2}{\kappa^2 \Delta^2} (m^2 + n^2) } \right).
\label{WignerAS} 
\end{equation} 
Some limiting cases are of interest. Suppose that $\Delta \ll \kappa^{-1}, h$. In this case, 
the double sums in (\ref{WignerS}) and (\ref{WignerAS}) vanish and the 
case of uniformly charged planes is recovered. Since each of the terms being summed 
is positive, discreteness of charge always has the 
effect of  increasing the magnitude of the interaction. The dependence of the disjoining pressure 
on $\kappa h$ and $\kappa \Delta$ is shown in Figure~\ref{fig:wigner}.

\subsection{Two dimensional random distributions}
\label{subsec_Random}
We now consider two parallel infinite planes with charge 
distributions $\sigma_\pm=\sigma_{0\pm}+\sigma_{f\pm}$, where the
$\sigma_{f\pm}$ are random distributions with zero mean and finite
variance. As in 
Section \ref{sec_NormalForce}\ref{subsec_GenProb}, 
we construct the symmetric and antisymmetric 
combinations $\sigma_S = (\sigma_{+} + \sigma_{-})/2$ and $\sigma_A =
(\sigma_{+} - \sigma_{-})/2$. The disjoining pressure is then given by
(\ref{FzGen}) in terms of $\langle|\hat\sigma_S|^2\rangle$ and
$\langle|\hat\sigma_A|^2\rangle$, where $\langle \ \rangle$ indicates an average 
over a suitable statistical ensemble. The 
force is therefore the sum of that due to uniform symmetric and
antisymmetric charge distributions $\langle\sigma_S\rangle$ and
$\langle\sigma_A\rangle$, together
with contributions from the fluctuations. In the following sections we
therefore consider symmetric and antisymmetric charge distributions with zero mean and variance
$\langle\sigma_S^2\rangle$ or  $\langle\sigma_A^2\rangle$.

\subsubsection{Symmetric problem}
We first consider the symmetric problem characterized by a two
point autocorrelation 
\begin{equation}
{\cal C} (\rho) = \frac{\langle \sigma_{S} (x,y) \sigma_{S} (x^{\prime},y^{\prime})  \rangle}{
\langle \sigma_{S}^{2} \rangle},
\label{defineCalC}
\end{equation} 
where $\rho = [(x-x^{\prime})^{2} + (y-y^{\prime})^{2}]^{1/2}$. Our assumption that the 
system is homogeneous and isotropic cannot hold 
if the charged regions are of finite lateral extent. Thus, we must assume that the characteristic size $L$
of the charged domain $D_{0}$ is much larger than the correlation
length $\alpha^{-1}$
characterizing the distance over which the function ${\cal C} (\rho)$ decays to zero, and eventually 
pass to the limit $\alpha L \rightarrow \infty$.
Equation~(\ref{Fzsym}) can be written in terms of the ensemble average as 
\begin{equation}
F_z = \frac{2 \pi^2}{\epsilon} \int \langle |\hat{\sigma}_{S}|^{2} \rangle \cosech^{2} (Kh)  \; dk_x dk_y,
\label{Fzsym1}
\end{equation}
and $\langle |\hat{\sigma}_{S}|^{2} \rangle$ may be expressed in terms of the two point 
correlation as 
\begin{equation}
\langle |\hat{\sigma}_{S}|^{2} \rangle  = \frac{1}{(2 \pi)^{4}} \int_{D_0} dx \, dy \int_{D_0} dx^{\prime} \, dy^{\prime} 
\langle \sigma_{S} (x,y) \sigma_{S} (x^{\prime},y^{\prime}) \rangle 
\exp( - i \bm{k} \cdot \bm{\rho} ),
\label{sigma_S_squared_two_pt_correlation}
\end{equation}
where $\bm{\rho}$ is the vector that points from $(x,y)$ to $(x^{\prime},y^{\prime})$.
Introducing the two point correlation (\ref{defineCalC}) into
(\ref{sigma_S_squared_two_pt_correlation}) we obtain, from (\ref{Fzsym1}),
\begin{equation}
\Pi = \frac{F_{z}}{A}  = \frac{\langle \sigma_{S}^{2} \rangle}{2 \epsilon} \frac{1}{(2 \pi)^2} 
\int  \left\{
 \int {\cal C} (\rho ) \exp( - i \bm{k} \cdot \bm{\rho} ) 
\, d \bm{\rho} \right\}  \cosech^{2} (Kh)  \; dk_x dk_y,
\end{equation}
where $A$ is the area of the charged domain $D_0$, and the integration with respect 
to $\bm{\rho}$ extends 
over the infinite $(x,y)$ plane. The latter integral  can be split into angular 
and radial components, with the angular integral expressible in terms of the Bessel function 
$J_{0} ( k \rho)$. Thus, 
\begin{equation}
\frac{\Pi}{\langle \sigma_{S}^{2} \rangle / (2 \epsilon)} =  
\int_{0}^{\infty}   \left\{  \int_{0}^{\infty} \rho \; {\cal C} (\rho ) J_{0} (k \rho) \; d \rho \right\}
k \cosech^{2} (Kh)  \; dk.
\label{PiRand}
\end{equation}
If we express the two point correlation as ${\cal C} (\rho) = {\cal C}_{0} (\alpha \rho)$ where 
${\cal C}_{0}$ is parameter free and $\alpha^{-1}$ is the correlation length, then (\ref{PiRand}), 
rewritten using  dimensionless variables $\bar{\rho} =\alpha \rho$ and
$\bar{k} = k / \alpha$, becomes
\begin{equation}
\frac{\Pi}{\langle \sigma_{S}^{2} \rangle / (2 \epsilon)} =  
\int_{0}^{\infty}   \left\{  \int_{0}^{\infty} \bar{\rho} \; {\cal C}_{0} (\bar{\rho} ) J_{0} (\bar{\rho} \bar{k} ) \; d \bar{\rho} \right\}
\bar{k} \cosech^{2} \left( \kappa h \sqrt{1 + \frac{\alpha^2}{\kappa^2}  \bar{k}^{2} } \right) \; d \bar{k}.
\label{PiRand1}
\end{equation}
Thus, the dimensionless disjoining pressure is a function of the two independent dimensionless 
parameters $\kappa h$ and $\alpha / \kappa$. Two special limits are of interest.
\paragraph{Long correlation length}
Let us consider the situation where the 
correlation length greatly exceeds the Debye length, i.e. $\alpha^{-1} \gg \kappa^{-1}$.
The term $\alpha^2 / \kappa^2$ in (\ref{PiRand1}) may now be
neglected, leaving
\begin{equation}
\frac{\Pi}{\langle \sigma_{S}^{2} \rangle / (2 \epsilon)} =  
\cosech^{2} ( \kappa h )
\int_{0}^{\infty}   \left\{  \int_{0}^{\infty} \bar{\rho} \; {\cal C}_{0} (\bar{\rho} ) J_{0} (\bar{\rho} \bar{k} ) \; d \bar{\rho} \right\}
\bar{k}  \; d \bar{k}.
\label{PiRandLong}
\end{equation}
The double integral on the right hand side of the equation evaluates
to ${\cal C}_0(0)=1$. To see this, we either appeal to the Hankel
transform pair $(1, \delta(\bar\rho)/\bar\rho)$, or we interchange 
the order of integration after introducing a regularizing factor $\exp(- a \bar{k})$ in the integrand. 
The integral with respect to $\bar{k}$ can be evaluated exactly, 
\begin{equation} 
\int_{0}^{\infty} \bar{k} e^{- a \bar{k}  } J_{0} ( \bar{\rho} \bar{k} ) \, d \bar{k} 
 = \frac{a}{( a^2 + \bar{\rho}^2)^{3/2}},
 \label{Intwrtk}
\end{equation}
by parametric differentiation with respect to $a$ of the standard integral  \cite{abramowitz_handbook_1964} 
\begin{equation} 
\int_{0}^{\infty} e^{- a x } J_{0} ( b x ) \, dx = \frac{1}{\sqrt{a^2 + b^2}}.
\end{equation} 
On substituting (\ref{Intwrtk}) in (\ref{PiRandLong}) and changing variables to $\xi = \bar{\rho}/a$,
 the integral becomes 
 \begin{equation} 
\int_{0}^{\infty} \frac{\xi {\cal C}_{0} ( a \xi )\; d \xi}{(1 + \xi^2)^{3/2}} 
\rightarrow {\cal C}_{0} ( 0 ) \int_{0}^{\infty} \frac{\xi\; d \xi  }{(1 +
  \xi^2)^{3/2}} 
= {\cal C}_{0} (0)=1,
\end{equation} 
on taking the limit $a \rightarrow 0$.
Thus, in the limit of long correlation length,
\begin{equation}
\frac{\Pi}{\langle \sigma_{S}^{2} \rangle / (2 \epsilon)} =  
\cosech^{2} ( \kappa h ),
\label{PiRandLong1}
\end{equation}
the same as (\ref{PiSymInf}) 
for parallel planes with uniform charge 
$\langle \sigma_{S}^{2} \rangle^{1/2}$ on each surface, and independent 
of the functional form of the two point correlation function. 
\paragraph{Short correlation length}
When the fluctuations in the charge are very fine grained, with
$\alpha^{-1} \ll \kappa^{-1}$, we introduce the small parameter $\delta = \kappa / \alpha$, 
and the rescaled variable $\eta = \bar{k} / \delta=k/\kappa$ in (\ref{PiRand1}):
\begin{equation}
\frac{\Pi}{\langle \sigma_{S}^{2} \rangle / (2 \epsilon)} =  
\delta^2 \int_{0}^{\infty}   \left\{  \int_{0}^{\infty} \bar{\rho} \; {\cal C}_{0} (\bar{\rho} ) J_{0} (\delta \bar{\rho} \eta )
 \; d \bar{\rho} \right\}
\eta  \cosech^{2} \left( \kappa h \sqrt{1 + \eta^{2} } \right) \; d \eta.
\label{PiRandShort}
\end{equation}
In the limit $\delta \rightarrow 0$, $J_{0} (\delta \bar{\rho} \eta) \rightarrow 1$ and therefore
\begin{equation}
\frac{\Pi}{\langle \sigma_{S}^{2} \rangle / (2 \epsilon)} =   \frac{\beta \kappa^2}{ \alpha^2}  {\cal F} (\kappa h),
\label{PiRandShort1}
\end{equation} 
where 
\begin{equation} 
 \beta = \frac{1}{2} \int_{0}^{\infty} \bar{\rho} \, {\cal C}_{0}  (\bar{\rho}) \; d \bar{\rho} 
 \label{definec}
 \end{equation} 
 is a constant of order unity, and
the function ${\cal F} (x)$ is defined as 
\begin{align}
{\cal F} (x) &=  \int_{0}^{\infty} \frac{2 \eta \, d \eta }{\sinh^{2} ( x \sqrt{1 + 
 \eta^2 } )} 
 = \frac{2}{x^2}\int_{x}^{\infty} \frac{t\,dt}{\sinh^{2} t }
=\frac{2}{x^2}\left\lbrack x\coth x-\ln(\sinh x)-\ln 2\right\rbrack,
 \label{defineF}
\\
&=\frac{2}{x^2}\left\lbrack\ln (x^{-1})+1-\ln
  2+\cdots\right\rbrack,\hskip 20pt 0<x\ll 1,
\label{smallxRigor}
\\
&\sim \frac{4}{x} e^{-2x},\hskip 20pt x\gg 1.
 \label{Fxlarge}
\end{align}
 Thus, the disjoining pressure (\ref{PiRandShort1}) decreases as
 $\alpha^{-2}$ as $\alpha\rightarrow\infty$ (i.e. as the correlation
 length $\alpha^{-1}$ becomes smaller).

  \begin{figure}[t]
    \centering
    \includegraphics[width=0.48 \textwidth]{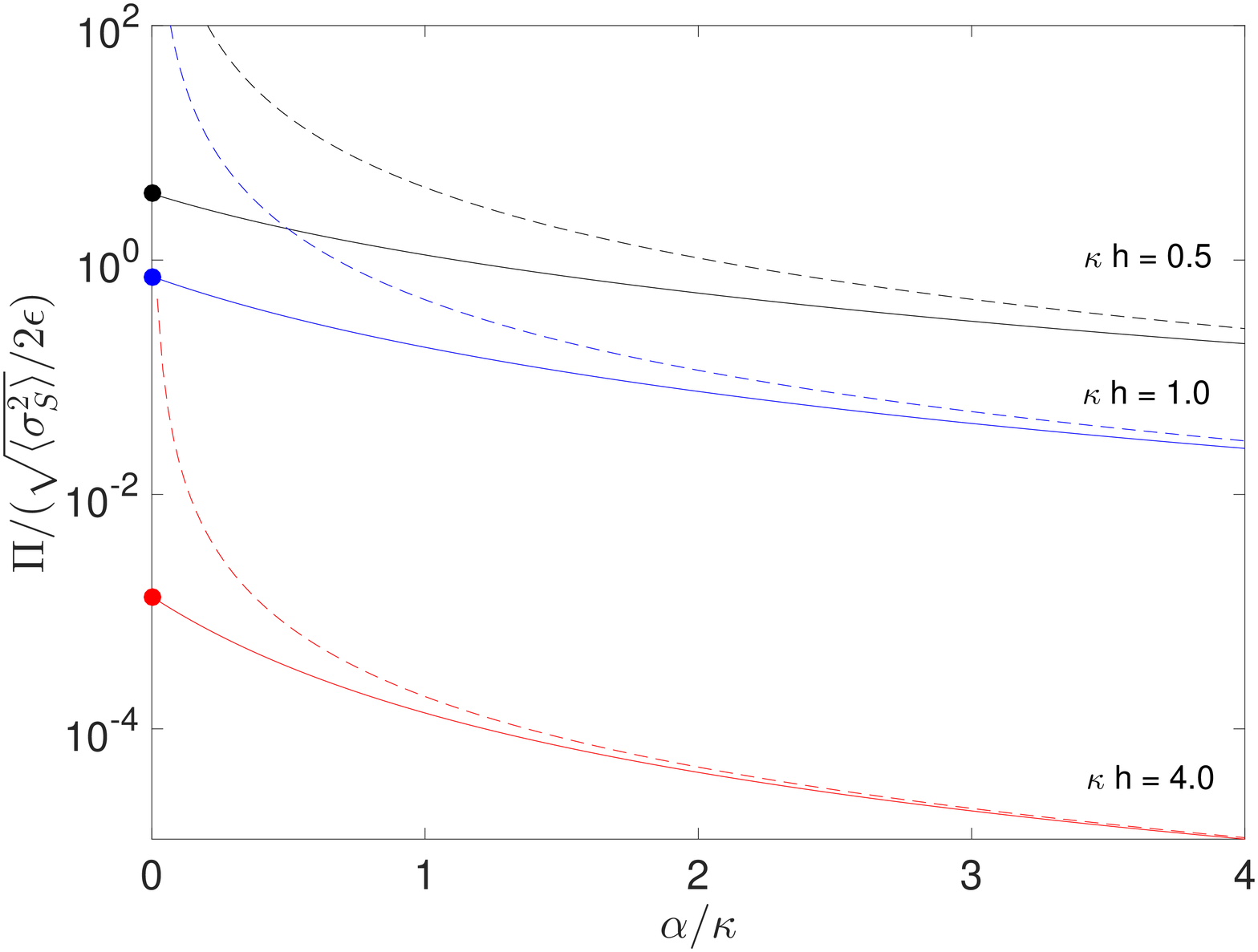}
    \includegraphics[width=0.48 \textwidth]{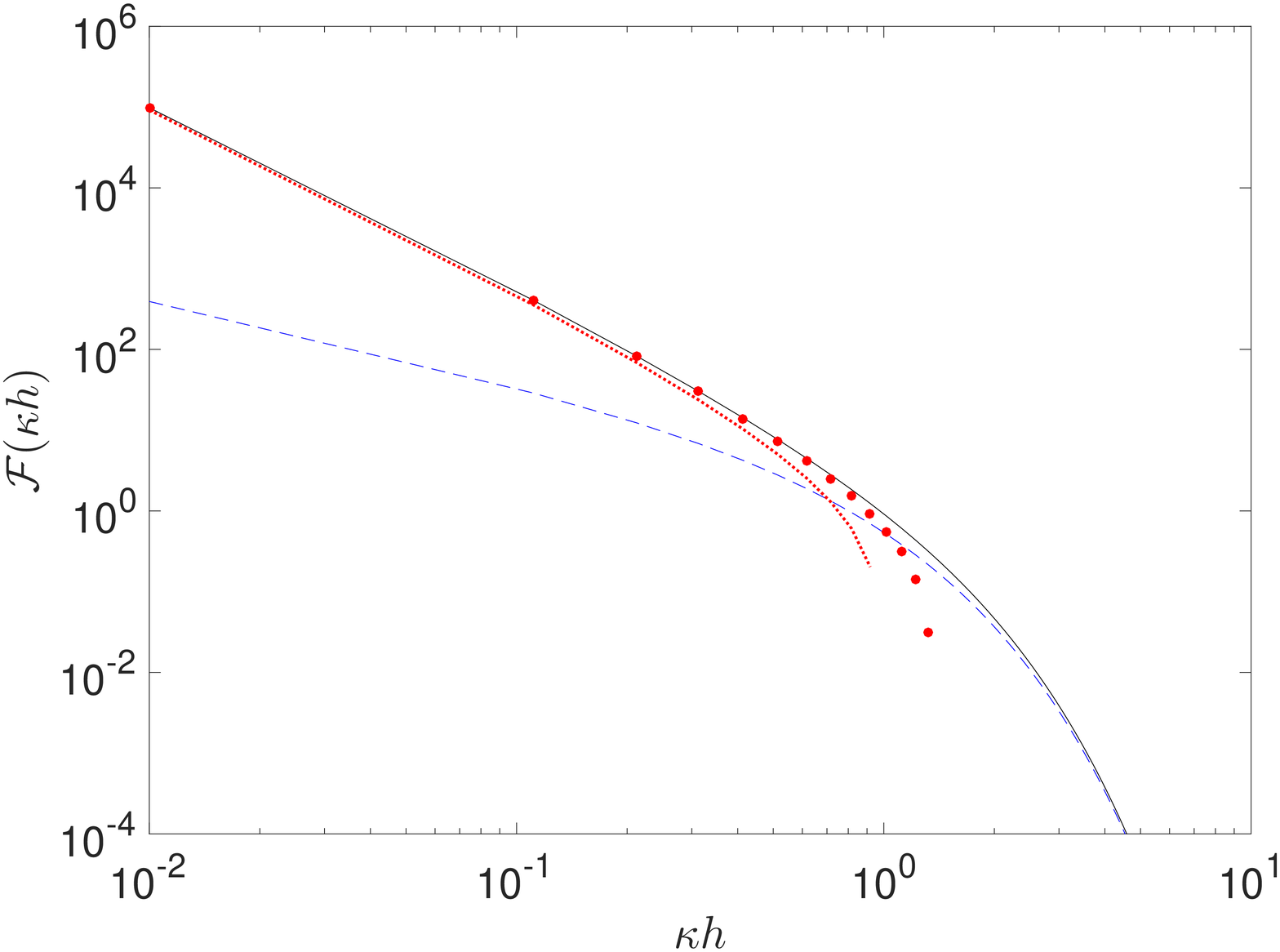} 
    \caption{{\em Left panel}: The normalized disjoining pressure from
      (\ref{PiRandExp}) for the autocorrelation (\ref{CExp}), for 
    fixed values of $\kappa h$ as a function of $\alpha / \kappa$ (solid lines).
    The corresponding dashed lines are the short correlation length ($\alpha / \kappa \gg 1)$ 
    asymptotic approximation  (\ref{PiRandShort1}). The filled circles are the 
    long correlation length limit ($\alpha / \kappa \rightarrow 0$) (\ref{PiRandLong1}).
    {\em Right panel}: The function ${\cal F} (\kappa h)$ given by (\ref{defineF}) 
   (solid line). Also shown are the large argument approximation (\ref{Fxlarge}) (dashed line) 
   and the one (dotted line) and two term (filled circles) small argument 
   asymptotic approximation (\ref{smallxRigor}).}
    \label{fig:dprnd_calF}
 \end{figure}

\paragraph{Intermediate correlation length}
If the correlation length $\alpha^{-1}$ is neither long nor short
compared to the Debye length $\kappa^{-1}$, we determine the disjoining pressure by evaluating 
the integral in (\ref{PiRand1}).
As an example, we consider a two point correlation function of the form
\begin{equation} 
{\cal C} (\rho) = \exp( - \alpha \rho ).
\label{CExp}
\end{equation}
Thus, ${\cal C}_{0} (\bar{\rho} ) = \exp (- \bar{\rho} )$ and from (\ref{definec}), $\beta = 1/2$.
The integral with respect to $\bar{\rho}$ in (\ref{PiRand1}) may be obtained in closed form using 
(\ref{Intwrtk}), thus
\begin{equation}
\frac{\Pi}{\langle \sigma_{S}^{2} \rangle / (2 \epsilon)} =  
\int_{0}^{\infty} \frac{\bar{k}}{(1+\bar{k}^2)^{3/2}} \cosech^{2} \left( \kappa h \sqrt{1 + \frac{\alpha^2}{\kappa^2}  \bar{k}^{2} } \right) \; d \bar{k}.
\label{PiRandExp}
\end{equation}
The integral in (\ref{PiRandExp}) was evaluated numerically and Figure~\ref{fig:dprnd_calF}
shows the resulting normalized disjoining pressure as a function of $\alpha / \kappa$. Also shown 
are the asymptotic limits at long correlation lengths, (\ref{PiRandLong1}), and short 
correlation lengths, (\ref{PiRandShort1}). 
\subsubsection{Antisymmetric problem}
In the case of an antisymmetric charge distribution (\ref{PiRand}) is replaced by 
\begin{equation}
\frac{\Pi}{\langle \sigma_{A}^{2} \rangle / (2 \epsilon)} =  
- \int_{0}^{\infty}   \left\{  \int_{0}^{\infty} \rho \; {\cal C} (\rho ) J_{0} (k \rho) \; d \rho \right\}
k \sech^{2} (Kh)  \; dk.
\label{PiRandA}
\end{equation}
At long correlation lengths we get the classical result (\ref{eq4PiLinf1}) for uniform plates with charge 
density $\langle \sigma_{A}^{2} \rangle^{1/2}$, i.e.
$\Pi = - ( \langle \sigma_{A}^{2} \rangle / 2 \epsilon ) \sech^{2} ( \kappa h )$,
and at short correlation lengths 
\begin{equation}
\frac{\Pi}{\langle \sigma_{A}^{2} \rangle / (2 \epsilon)} 
= - \frac{\beta \kappa^2}{\alpha^2} {\cal G} (\kappa h),
\end{equation}
where
\begin{equation}
{\cal G} (x) = \frac{2}{x^2}\int_{x}^{\infty} \frac{t\,dt}{\cosh^{2} t }
=\frac{2}{x^2}\left\lbrack \ln 2-x\tanh x+\ln\cosh(x)\right\rbrack.
\label{defineG}
\end{equation} 
When $x$ is large, 
${\cal G} (x) \sim {\cal F} (x) \sim  (4/x) \exp(-2x)$. 
When $x$ is small, ${\cal G} (x) \sim 2\ln(2)/x^{2}$.
For the special choice (\ref{CExp}) of the two point correlation function, the dimensionless 
disjoining pressure in the antisymmetric problem becomes 
\begin{equation}
\frac{\Pi}{\langle \sigma_{A}^{2} \rangle / (2 \epsilon)} =  
- \int_{0}^{\infty} \frac{\bar{k}}{(1+\bar{k}^2)^{3/2}} \sech^{2} \left( \kappa h \sqrt{1 + \frac{\alpha^2}{\kappa^2}  \bar{k}^{2} } \right) \; d \bar{k}
\label{PiRandExpA}
\end{equation}
which may be compared with the result (\ref{PiRandExp}) for the
symmetric case.

\subsubsection{General problem}
By way of example, we consider two plates, each 
with zero net charge, and with no cross-correlation between the charge
distributions on the two plates. Thus, 
$\langle \sigma_{+} \rangle = \langle \sigma_{-} \rangle = \langle \sigma_{+} \sigma_{-}^{\prime} \rangle = 0$,
where the prime indicates evaluation at $(x^{\prime},y^{\prime})$ rather than 
$(x,y)$. Let us also suppose that the mean squared charge density is $\langle \sigma^2 \rangle$ 
and the autocorrelation is ${\cal C} (\rho)$ on either plate. Thus, 
$\langle \sigma_{+}^2 \rangle = \langle \sigma_{-}^2 \rangle = \langle \sigma^2 \rangle$ and 
$ \langle \sigma_{+} \sigma_{+}^{\prime} \rangle = 
\langle \sigma_{-} \sigma_{-}^{\prime} \rangle = \langle \sigma^2 \rangle {\cal C} (\rho)$.
It follows that 
$\langle \sigma_{S} \sigma_{S}^{\prime} \rangle = 
\langle \sigma_{A} \sigma_{A}^{\prime} \rangle = \langle \sigma^2 \rangle {\cal C} (\rho)/2$,
i.e. the autocorrelations of $\sigma_S$ and $\sigma_A$ are non-zero, even though the charge distributions on the two plates are uncorrelated.
Thus, on combining (\ref{PiRand}) and (\ref{PiRandA}) we have 
\begin{equation}
\frac{\Pi}{\langle \sigma^{2} \rangle / (2 \epsilon)} =  
\int_{0}^{\infty}   \left\{  \int_{0}^{\infty} \rho \; {\cal C} (\rho ) J_{0} (k \rho) \; d \rho \right\}
2 k \cosech^{2} (2Kh)  \; dk.
\label{PiRandany}
\end{equation}
Note that the disjoining pressure is non-zero even though the charge fluctuations 
on the two plates are uncorrelated with each other. This is because interactions between elements with 
like charge and unlike charge do not contribute equally to the total force.
Velegol and Thwar \cite{velegol_analytical_2001}  concluded that planes with random, uncorrelated charge attract each other, rather than repel. However, their analysis differs from ours in two respects. (i) They assume a force law at each point $(x,y)$ that is identical to that between planes with uniform charge. This neglects the tendency of the charge cloud to smooth itself out over the Debye length scale $\kappa^{-1}$, and is therefore unsatisfactory when the correlation length is small compared to the Debye length. (ii) More importantly, they assume that the potentials on the planes (rather than the surface charge densities) are held constant, leading to a different force 
law \cite{bell_calculation_1972}.

\section{Discussion}
\label{sec_Discussion}
We have presented explicit formulas for the disjoining pressure between inhomogeneously charged planes for various charge distributions that should be useful in applications.Though our starting point, 
(\ref{Sigmazz1}), is not restricted to low potentials, in the subsequent development we have adopted the Debye-Huckel linearized theory in order to derive expressions for the disjoining pressure. It is, however, well known that the Debye-Huckel linearized theory is inadequate when potentials on charged surfaces exceed the thermal scale $k_BT/e\approx 25\rm\ mV$. 
This is particularly relevant when the surfaces are very close together ($\kappa h \ll 1$),
as in this case Debye shielding is ineffective at reducing the potentials near the surfaces.

This nonlinear regime can be explored by perturbation methods or by 
numerical solution of the underlying Poisson-Boltzmann equation
\cite{white_increased_2002,leote_de_carvalho_non-linear_1998}. However, if the surface charge 
density is very large, the disjoining pressure can be obtained by a minor modification 
of the method presented here. By `very large' we mean that the surface 
charge density, $\sigma$, greatly exceeds the critical charge density, $\sigma_{c}= e / (\ell_{B} \kappa^{-1} )$ 
where $\ell_{B} = e^{2} / ( 4 \pi \epsilon k_{B} T )$
 is the Bjerrum length. This critical density corresponds to the presence of one electronic charge 
 for every square of side $h = (\ell_{B} / \kappa)^{1/2}$, the geometric mean of the Debye length and Bjerrum length. 
 When $\sigma \gg \sigma_c$,
 the Debye layer may be regarded as an inner layer where the potential is comparable to the 
 thermal scale, and an outer diffuse layer where the Debye-H\"{u}ckel approximation may be employed.
 The extent of this inner layer is defined by the length scale $\ell = \kappa^{-1} \ln (2 \pi z \sigma / \sigma_c)$,
 for a symmetric electrolyte with valence $z$. If $\sigma / \sigma_c$ is sufficiently large that 
 $\ell$ is much less than each of the other three length scales $\kappa^{-1}$, $h$ and $L$ 
 then the geometry of the nonlinear region may be regarded as planar. In this 
 situation, an exact solution of the Poisson Boltzmann equation is available for a binary symmetric 
 electrolyte and can be matched to the outer diffuse layer. This matching procedure is well known \cite{russel_colloidal_1989} and leads to the conclusion that in the 
 far field the classical linear Debye-H\"{u}ckel solution is valid provided one replaces the true 
 charge density, $\sigma$, by an ``effective'' or ``renormalized'' charge density
 $\sigma_{e} = (\sigma_c / \pi z) \ln ( 2 \pi z \sigma / \sigma_c )$.
Thus, all of the results presented in this paper may be used but with $\sigma$ replaced by 
the corresponding $\sigma_e$ at all points on the surface. 
  
 In this paper we have neglected any penetration of the electric fields into the dielectric substrate
 bounding the electrolyte. Since the relative permittivity of water is very large compared to that of common substrates (e.g. plastics, glass, lipids etc.), this neglect is usually a very good approximation except in special situations such as near dielectric-electrolyte 
 interfaces with sharp corners \cite{thamida_nonlinear_2002}.
 However, this assumption is not essential to the development presented here and can 
be avoided by replacing (\ref{BVprobSoln}) by 
a potential that is continuous at the interface between the electrolyte and the
dielectric substrate, with a jump in derivative corresponding to the
surface charge at the interface.
  The finite permittivity contrast between electrolyte and dielectric will cause 
 some flux leakage into the solid which will alter the calculation of the edge effect in a 
 non-trivial manner. Under appropriate conditions, this correction may be significant. We leave an investigation of this to future work.
 
\section{Conclusion}
\label{sec_conclusion}
We have presented a method for calculating the normal force between inhomogeneously 
charged planes bounding an electrolyte. The problem was considered within the linearized Debye-H\"{u}ckel 
approximation that is valid when the electric potentials everywhere are small compared to the thermal 
scale $k_{B} T / e$. This problem has been considered by various authors (see Sec.~\ref{sec_intro}) 
in different contexts starting with the work of Richmond \cite{richmond_electrical_1974,richmond_electrical_1975}. 
In all cases, the basic approach involved developing an expression for the free energy by the ``charging 
method'' discussed by Richmond. This  has the advantage that the free energy  is a scalar, and 
once it is known all force components can be determined by differentiation. We have presented here 
an alternate approach that may be much more efficient if one is only interested in the normal 
component of the interaction force (disjoining pressure). We have applied this method to the 
various classes of problems and in each case we have presented explicit 
formulas for the disjoining pressure that should be useful in applications. 

A distribution that appears not to have been considered before is that of a pair of infinite planes with 
a central charged section.
This configuration is interesting because it provides a simple model for understanding the effects of loss 
of lateral confinement of the Debye layer.
Previous work \cite{ghosal_repulsion_2016} considered Debye layer overspill
out of the gap between two dielectric blocks that face each other across a gap of width $2h$ and are of
lateral extent $2L$. An analysis was presented in the Debye-H\"{u}ckel 
limit for the case of Debye length ($\kappa^{-1}$) much greater than $h$.
This block geometry (`unconfined') and the case of infinite parallel planes (`confined') 
considered here may be regarded as special cases of a general one dimensional problem in which the gap width $2h(x)$ and charge densities
$\sigma = \sigma_{\pm} (x)$ are functions of position $x$. These special cases
have enabled us to demonstrate the effects of charge overspill in problems
that are sufficiently simple that analytic expressions can be obtained for
the consequent reduction in the force between the charged surfaces.
In conclusion, we point out that the ``confined'' problem 
studied here is similar in some ways to the problem of electro-osmotic flow past a 
step change in surface charge considered by Yariv~\cite{yariv_electro-osmotic_2004}.



\section*{Acknowledgement} 
We thank an anonymous referee for suggesting improvements
to Sec. \ref{sec_Applications}\ref{subsec_Random}.
J.D.S. thanks the Department of Applied Mathematics
\& Theoretical Physics, University of Cambridge, for hospitality.
\section*{Appendix A: Derivation of (\ref{eq4Pismallh}) and (\ref{PB_lostlength_maintext})}
Using $\bar{\phi}(x)$ to denote the average of the potential $\phi(x,z)$ over the channel cross-section,
$z$, we have, from (\ref{DHeq}) and (\ref{DHeqBC}), 
$\partial_{xx} \bar{\phi} - \kappa^2 \bar{\phi} = - \sigma_0 / \epsilon h$  if  $|x| < L$ and zero 
otherwise. Solutions are symmetric about $x=0$ and take the form
\begin{equation}
 \bar{\phi} (x) =
  \begin{cases} 
\sigma_0 / (\epsilon \kappa^2 h)  + A \cosh (\kappa x) &  \text{if } |x| \leq L,\\
 B \exp(-\kappa |x|) & \text{if } |x| >  L.
 \end{cases}
\end{equation}
Continuity of  $\bar{\phi}$ and $\partial_{x} \bar{\phi}$ at $x=L$ determines the constants 
$A$ and $B$:
\begin{equation}
A = -\frac{\sigma_0 \exp(-\kappa L)}{\epsilon \kappa^2 h}, \qquad 
B = \frac{\sigma_0 \sinh(\kappa L)}{\epsilon \kappa^2 h}.
\label{B_evaluated}
\end{equation}
In the limit $\kappa h \ll 1$, $\phi(x,z) \sim \bar{\phi} (x) + O(\kappa^2 h^2)$, and thus,
from (\ref{Sigmazz1}),
\begin{equation}
 \Sigma_{zz} \sim  -\frac{\epsilon}{2}
  \begin{cases} 
 2 B^2  \kappa^2 \exp( -2\kappa |x| ) &  \text{if } |x| > L,\\
  \frac{\sigma_0^2}{\epsilon^2 \kappa^2 h^2}+A^2 \kappa^2 \cosh^2(\kappa x)  
  +\frac{2A \sigma_0}{\epsilon h} \cosh(\kappa x)  +A^2 \kappa^2 \sinh^2(\kappa x) 
  & \text{if } |x| \leq  L.
 \end{cases}
\label{sol4phibar}
\end{equation}
Hence, the total force is
\begin{equation} 
F = -2 \int_0^\infty \Sigma_{zz} \, dx 
= \frac{\sigma_0^2 L}{\epsilon \kappa^2 h^2} \left[1 - \frac{1- \exp( -2\kappa L )}{2 \kappa L} \right]
\end{equation}
and on dividing by $2L$,  (\ref{eq4Pismallh}) follows.

The total charge in $x>L$ is
\begin{equation}
Q = -2h\epsilon\kappa^2  \int_L^\infty \overline{\phi} \, dx
= -2h\epsilon\kappa B\text{e}^{-\kappa L} 
= -\frac{2\sigma_0}{\kappa[1+\coth(\kappa L)]}.
\label{eqQ}
\end{equation}
This is the amount of charge needed to neutralize a
length $\Lambda=-Q/(2\sigma_0)$ along the interface, the ``lost length'' 
\cite{ghosal_repulsion_2016}. Using  (\ref{eqQ}) for $Q$, (\ref{PB_lostlength_maintext}) follows.

Equation ({\ref{eq4Pismallh}) may also be derived by 
contour integration (in the complex plane) of the integral (\ref{eq4PiA}), which
is the real part of
\begin{equation}
I = \frac{1}{2 \pi p} \int_{-\infty}^{+\infty} \frac{1 - \exp(2ipz)}{z^{2} (1 + z^{2})} \; dz,
\end{equation} 
where $p = \kappa L$.
We consider a closed contour 
$C=C_{0} \cup C_{\varepsilon} \cup C_{R}$, where
$C_{0} = (-\infty,-\varepsilon) \cup (\varepsilon,\infty)$, $C_{\varepsilon}$ is the semicircle with center 
at the origin and radius $\varepsilon>0$ that lies in the upper half $\Im (z) > 0$ of the complex plane, and
$C_{R}$ is the semicircle with center at the origin and radius $R\gg 1$, again
in the upper half $\Im (z) > 0$ of the complex plane. We take the limit 
$\varepsilon \rightarrow 0$ and $R \rightarrow \infty$.  Denoting by $I_{0}$, $I_{\varepsilon}$ 
and $I_{R}$ the contributions to the integral from the parts of the domain $C_{0}$, 
$C_{\varepsilon}$ and $C_{R}$ and evaluating the residue from the pole at $z=i$,
we find $I_{0} + I_{\varepsilon} + I_{R} = (\exp(-2p) - 1)/(2p)$. 
Clearly, $I_{R} \rightarrow 0$ as $R \rightarrow \infty$ and $I_{\varepsilon} \rightarrow -1$ as 
$\varepsilon \rightarrow 0$. Thus, 
\begin{equation} 
I = 1 - \frac{1}{2p} \left[ 1 - \exp(-2p) \right],
\end{equation} 
and (\ref{eq4Pismallh}) follows.

\end{document}